\documentclass[12pt, prd, aps, preprintnumbers, nofootinbib]{revtex4}

\usepackage{epsfig}
\usepackage{amsmath}
\usepackage{amssymb}
\usepackage{psfrag}

\newcommand{\ie}{{\em i.e.~}}
\newcommand{\eg}{{\em e.g.~}}
\newcommand{\cf}{\emph{cf.~}}

\newcommand{\be}{\begin{equation}}
\newcommand{\ee}{\end{equation}}
\newcommand{\bee}{\begin{equation*}}
\newcommand{\eee}{\end{equation*}}
\newcommand{\bea}{\begin{eqnarray}}
\newcommand{\eea}{\end{eqnarray}}
\newcommand{\bean}{\begin{eqnarray*}}
\newcommand{\eean}{\end{eqnarray*}}

\newcommand{\nn}{\nonumber}

\newcommand{\lp}{\left(}
\newcommand{\rp}{\right)}

\begin{document}

\preprint{CERN-PH-TH/2010-151, UAB-FT-683, ULB-TH/10-19}

\title{Gravitational backreaction effects on \\ the holographic phase transition}

\author{T.~Konstandin}
\email{tkonstan@cern.ch}
\affiliation{
CERN Physics Department, Theory Division, CH-1211
Geneva 23, Switzerland
}

\author{G.~Nardini}
\email{germano.nardini@ulb.ac.be}
\affiliation{
Service de Physique Th\'eorique,
Universit\'e Libre de Bruxelles, 1050 Brussels, Belgium
}

\author{M.~Quiros}
\email{quiros@ifae.es}
\affiliation{
ICREA, Instituci\`o Catalana de Recerca i Estudis Avan\c{c}ats,
IFAE \\ Universitat Aut{\`o}noma de Barcelona,
08193 Bellaterra, Barcelona, Spain
}

\date{\today}

\begin{abstract}
We study radion stabilization in the compact Randall-Sundrum model by
introducing a bulk scalar field, as in the Goldberger and Wise
mechanism, but (partially) taking into account the backreactions from
the scalar field on the metric. Our generalization reconciles the
radion potential found by Goldberger and Wise with the radion mass
obtained with the so-called superpotential method where backreaction
is fully considered. Moreover we study the holographic phase
transition and its gravitational wave signals in this model. The
improved control over backreactions opens up a large region in
parameter space and leads, compared to former analysis, to weaker
constraints on the rank $N$ of the dual gauge theory. We conclude
that, in the regime where the $1/N$ expansion is justified, the
gravitational wave signal is detectable by LISA.
\end{abstract}

\maketitle

\section{Introduction}
Even though the Standard Model (SM) of particle physics has achieved
many impressive experimental successes it fails to provide an
explanation to some experimental and theoretical issues. On the
experimental side, neither the observed dark matter density can be
explained (the SM lacking an appropriate natural candidate for it) nor
the baryon asymmetry of the Universe can be accounted of, mainly
because the SM does not provide sizeable CP violation sources and a
strong enough first-order electroweak phase transition. On the
theoretical side there is no plausible explanation for the huge
hierarchy between the electroweak scale, responsible for the mass of
the weak gauge bosons, and the Planck scale, apparent in the weakness
of gravitational interactions.  Thus an extension of the SM seems
necessary.

Guided by the naturalness criterion to solve the Higgs hierarchy
problem one very attractive possibility is Randall-Sundrum (RS)
models, which are based on the framework of a compact warped extra
dimension~\cite{RS} with two branes localized on it: an ultraviolet
(UV) brane, which provides the UV cutoff of the higher dimensional
theory, and an infrared (IR) one, which spontaneously breaks the
conformal symmetry of the theory. In this class of models the
four-dimensional part of the metric has a strong dependence on the
fifth coordinate, which is unobservable macroscopically, and the
hierarchy between the two energy scales is generated by localizing the
relevant physical degrees of freedom responsible for electroweak
breaking at (or near) the IR brane. In this way the huge hierarchy can
be explained by a natural distance between the two branes of $\mathcal
O(10)$ times the fundamental five-dimensional Planck length.

To complete the picture the distance between the branes should not be
considered as a fundamental input but instead it should arise from a
stabilization mechanism. In fact this is even essential to avoid a
massless radion generating an unobserved fifth force, and to obtain
the observed Friedman-Robertson-Waker cosmology at late
times~\cite{Cline:1999ts, Csaki:1999mp}.
One elegant possibility to stabilize the brane distance is to assume
the existence of a bulk scalar with a five-dimensional
mass~\footnote{Alternatively, Casimir effect can also lead to a
  stabilizing potential~\cite{Garriga:2000jb, Garriga:2002vf}.} that
is slightly smaller than the fundamental Planck mass scale, as
pioneered by Goldberger and Wise~\cite{Goldberger:1999uk}. In this
setup one usually assumes the weak field limit in which case the
dynamics of the bulk scalar field and the metric decouple. This does
not only have the advantage of simplifying the analysis but also
facilitates the interpretation of the scalar action as a radion
potential.

In the modern context of the AdS/CFT correspondence~\cite{AdSCFT} this
kind of model can be interpreted as dual to a strongly coupled gauge
theory that for instance might serve as a UV
completion~\cite{Contino:2003ve} to little Higgs
models~\cite{ArkaniHamed:2001nc}. Usually the five-dimensional system
is considered in the limit of a large ratio between the Planck mass
and the bulk cosmological constant, which implies a large number of
degrees of freedom in the dual field theory and allows to neglect
stringy effects. However there is a certain tension between this
assumption and a viable cosmology: even though the model has the above
discussed brane setup at low temperature (called RS-GW in the
following), at high temperature it is represented by an
AdS-Schwarzschild (AdS-S) bulk metric according to the AdS/CFT
correspondence. Even though a first-order phase transition could allow
the Universe to escape from the (conformal) AdS-S to today's
(quasi-conformal) RS-GW phase~\cite{Creminelli:2001th}, its completion
leads to a stringent constraint on the above
ratio~\cite{Creminelli:2001th, Randall:2006py, Kaplan:2006yi,
Nardini:2007me, Reece:2010xj} which jeopardizes the consistency of the
original assumptions.

On the other hand several appealing features characterize this
necessary phase transition.  One of them is its extreme supercooling,
which may couple the conformal phase transition to the electroweak one
making the latter strong enough to plausibly explain the baryon
asymmetry of the Universe~\cite{Nardini:2007me}. Besides a further
feature is its potential testability: the phase transition could mark
the gravitational wave spectrum which could allow to probe most of the
interesting parameter space of the model~\cite{Randall:2006py}.

In conclusion given the success of RS models as extensions of the SM
we believe it is of importance to clarify the impact of the
phase transition on the validity of the models. This paper is
dedicated to this issue in the case of RS-GW models. Our aim is to
alleviate the tension between the ratio of the Planck mass to the
bulk cosmological constant and the phase transition completion and to
show how this scenario can be tested in the gravitational
wave spectrum.

In this paper we will proceed as follows.  After introducing some
notations and conventions in sec.~\ref{sec:notation} we will review in
sec.~\ref{sec:GW} the usual procedure to determine the effect of the
bulk scalar on the brane distance~\cite{Goldberger:1999uk,
Csaki:2000zn}. We will remark that the method of
ref.~\cite{Goldberger:1999uk} determines the effective potential of
the radion although without taking into account the backreactions of
the scalar field on the gravitational metric. On the contrary the
superpotential method~\cite{Csaki:2000zn} solves the problem exactly
but it does not provide the radion effective potential which is needed
to study the phase transition. For this reason in
sec.~\ref{sec:bk-react} we will present an alternative approach to
determine the radion potential in the regime of detuned brane tensions
and sizeable backreactions. It is based on fitting the radion
potential information that we can determine in this regime: the
position of the radion potential extrema and the radion mass and
cosmological constant in these extrema. The obtained potential will be
used in sec.~\ref{sec:pt} to study the phase transition and we will
find relaxed bounds that may alleviate the above problematic parameter
tension. The corresponding gravitational wave spectrum will be
determined in sec.~\ref{sec:gwo} and we will conclude by commenting on
the prospects of detection with forthcoming experiments as LISA.
Finally we will devote sec.~\ref{sec:concl} to summarize the main
results of the paper and we leave in appendix~\ref{sec:appx} some
technical details that we need in sec.~\ref{sec:bk-react} to determine
the radion mass in the presence of a cosmological constant.

\section{Notation and conventions\label{sec:notation}}
A very interesting feature of RS models with a scalar in the bulk is
that the scalar field can stabilize the brane
distance~\cite{Goldberger:1999uk, Csaki:2000zn}. The corresponding
five-dimensional (5D) action is given by
\be
\label{eq:def_action}
S = \int d^5x \sqrt{| \det g_{MN} |} \left[ - M^3 R + \frac12 
(\partial \phi)^2 - V(\phi) \right] 
- \sum_\alpha \int_{B_\alpha} d^4 x \sqrt{| \det \bar g_{\mu\nu} |} 
\lambda_\alpha (\phi)~,
\ee
where $M$ is the 5D Planck scale, $\lambda_\alpha$ and $V$ are the
brane and bulk potentials of the scalar field $\phi$, and the metric
$g_{MN}$ is defined in proper coordinates by
\bea
\label{eq:def_g}
ds^2 &=& e^{2 A(r)} \bar g_{\mu\nu} dx^\mu dx^\nu - dr^2~,\nonumber\\
\mathcal{M}_4: \hspace{1.68cm}\bar g_{\mu\nu}&=&\eta_{\mu\nu}~,\\
dS_4: \quad
\bar g_{\mu\nu} dx^\mu dx^\nu &=& dt^2- 
e^{2 \sqrt{\Lambda} t} (dx_1^2 + dx_2^2 + dx_3^2 )~, \\
AdS_4: \quad
\bar g_{\mu\nu} dx^\mu dx^\nu &=& -dx_3^2 -  
e^{-2 \sqrt{-\Lambda} x_3} (dx_1^2 + dx_2^2 - dt^2 )~.
\eea
Accordingly the induced four-dimensional (4D) metric $\bar g_{\mu\nu}$
is Minkowski, de Sitter or anti-de Sitter and the corresponding 4D
cosmological constant $\Lambda$ has mass dimension equal to 2. In all
cases the Ricci-tensor and scalar turn out to be
\bea
R_{\mu\nu} &=& e^{2 A} (4 A'^2 - A'' - 3 \Lambda e^{2 A}) \bar g_{\mu\nu}~,\\
R_{55} &=& - 4 A'^2 - 4  A''~, \\
R &=& 20 A'^2 + 8 A'' - 12 \Lambda e^{-2 A} \label{eq:R_scalar}~,
\eea
where $^\prime\equiv d/dr$. The equation of motion for the scalar field reads
\be
\label{eq:phi_eom}
\phi'' + 4 A' \phi' = \frac{\partial V}{\partial \phi}
+ \sum_\alpha \frac{\partial \lambda_\alpha}{\partial \phi} 
\delta(r-r_\alpha)~,
\ee
and the Einstein equations have the form
\bea
\label{eq:Einstein}
A'' + \Lambda e^{-2 A} &=& -\frac{\kappa^2}3 \phi'^2
-\frac{\kappa^2}3 \sum_\alpha \lambda_\alpha \delta(r-r_\alpha)~, \\
\label{eq:Einstein2}
A'^2 - \Lambda e^{-2 A} &=& -\frac{\kappa^2}6 V + \frac{\kappa^2}{12} \phi'^2~,
\eea
with $\kappa^2 = 1/(2M^3)$. 
Hereby the localized terms impose the following constraints (assuming
a $\mathbb{Z}_2$ symmetry across the branes)
\be
\label{eq:brane_constraints}
\left. A' \right|_{r_\alpha - \epsilon}^{r_\alpha + \epsilon} =
-\frac{\kappa^2}3 \lambda_\alpha(\phi(r_\alpha))~,~\quad
\left.  \phi' \right|_{r_\alpha - \epsilon}^{r_\alpha + \epsilon} = 
 \frac{\partial \lambda_\alpha(\phi(r_\alpha))}{\partial \phi}~.
\ee
Using these equations in the action one obtains 
\be
\label{eq:action_Lambda}
S = 6 M^3 \int d^5x \sqrt{| \det g_{MN} |} 
  e^{-2A} \Lambda~.
\ee

We would like to emphasize that this relation does not rely on any
approximation so far. The key observation is that from an effective
four-dimensional point of view the expansion parameter $\Lambda$ has
to be related to the value of the radion potential. If the system
(\ref{eq:phi_eom})-(\ref{eq:Einstein2}) allows for several solutions,
the corresponding values of $\Lambda$ allow to determine the difference in
potential energy of the radion between different configurations. Here we
will focus on positive cosmological constants and our
aim will be to determine $\Lambda$ with an accuracy that goes beyond the
usual weak field assumption $\phi^2 \ll M^3$~\cite{Goldberger:1999uk}.

\section{Stabilization and GW mechanism\label{sec:GW}}

Solving the system (\ref{eq:phi_eom})-(\ref{eq:Einstein2}) is for
generic scalar potentials a hard task. A possibility to overcome this
difficulty is to consider the quadratic Goldberger-Wise (GW)
scalar potential~\cite{Goldberger:1999uk}
\bea
\label{eq:GW_pot}
V(\phi) &=& -\frac{12M^3}{l^2} + \frac12 m^2 \phi^2~,\\
\label{eq:brane_pot}
\lambda_\alpha(\phi) &=& \lambda^0_\alpha +
\gamma_\alpha (\phi - v_\alpha)^2 
\qquad \textrm{with} 
\quad \gamma_\alpha \to \infty~,
\eea
where $l$ is the AdS length such that $1/l = k$ is of the order the Planck
scale, $\phi$ and $v_\alpha$ have mass dimension $3/2$ and
$\gamma_\alpha$ has mass dimension $1$. We assume that these
potentials are chosen such that the warping in the metric is close to
AdS, meaning that $A(r)$ in (\ref{eq:Einstein2}) is dominated by the
bulk vacuum energy
\be
\label{eq:constraints}
l^2 \phi^{\prime2}\ll 24 M^3, \quad 
l^2 m^2 \phi^2 \ll 24 M^3, \quad 
\Lambda \ll l^{-2}~. 
\ee
At leading order in $\phi$ the scalar sector then decouples
completely from gravity. The equations of motion for the metric are then
given by $\Lambda=0$ and $A' = -1/l$ and the gauge choice $A(0)=0$ leads
to the solution
\be
\label{eq:A_lead}
A(r) =- r/l~,
\ee
in the bulk. We assume the UV and IR branes to be located at $r_1 = 0$  and
$r_2=r_0$ respectively.

Finally under the assumptions (\ref{eq:constraints}) the solution of
the scalar field turns out to be
\bea
\label{eq:phi}
\phi(r) &=& v_1 e^{k_- r} + (v_2 - v_1 e^{k_- r_0}) 
\frac{e^{k_+ r} -  e^{k_- r}}
{ e^{k_+ r_0} -  e^{k_- r_0}}~,\\
\label{eq:kpm}
l k_\pm &=& 2 \pm \sqrt{4 + m^2 l^2}~,
\eea
where $v_\alpha$ denote the values of $\phi$ at the UV and IR
branes~\footnote{In the brane potentials~(\ref{eq:brane_pot}) the
linear terms satisfying eqs.~(\ref{eq:brane_constraints})
and~(\ref{eq:phi}) are omitted since the results of our analysis will
be independent of them.}.  We focus on negative values~\footnote{Most
results in our analysis are easily carried over to positive $k_-$.} of
$k_-$ and a solution of the hierarchy problem ($r_0\approx 37\,l$)
requires
\be
v_1 \sim v_2 \sim M^{3/2}~,\qquad
1 \sim e^{k_- r_0} \ll 10^{16} \sim e^{r_0 / l}~,
\ee
such that one obtains
\be
\label{eq:phiApprox}
\phi(r) \approx v_1 e^{k_- r} + ( v_2 - v_1 e^{k_- r_0} ) e^{k_+ (r-r_0)}~. 
\ee
Using this in the action (\ref{eq:def_action}) yields
\be
\label{eq:pot_GW}
 S   = l^{-1} \int d^4x \, \left\{  l k_- v_1^2  -  e^{-4 r_0/l}
 \Big[ (4 - l k_-  )( v_2  - v_1 e^{k_- r_0} )^2  
 + l k_-  v_2^2\Big] \right\}.
\ee
In ref.~\cite{Goldberger:1999uk} this is identified (after a
change of sign) with the potential of the radion $V_{GW}$ as
\be
S=-\int d^4x\,V_{GW}(r_0)~,
\ee
and has a minimum at $\xi=\xi_-$ (\ie~$r_0=r_-$) where
\be
\label{eq:ex_GW}
\xi\equiv \frac{v_1}{v_2} \, e^{k_- r_0}~,\qquad
\xi_- \equiv \frac{v_1}{v_2}e^{k_- r_-} = 
1 + |l k_-|^{1/2}/2 - l k_-/4 + O\left((l k_-)^{3/2}\right)\,.
\ee
The value of the potential at this minimum is of $\mathcal O(\text{TeV}^4)$ 
\be
\label{eq:pot_min_GW}
V_{GW}(r_-) - V_{GW}(\infty) \simeq l^{-1}
\, |l k_-|^{3/2} v_2^2 e^{-4 r_-/l}~. 
\ee

However this result is not conclusive. In an effective
description of the radion one would like to split the dynamics into
four-dimensional gravity and the radion degree of freedom. Hence the
radion potential should not only depend on the scalar part of the
action but might also receive a contribution from 5D
gravity. In the derivation of eq.~(\ref{eq:ex_GW}) we neglected
contributions to the field $A$ of order $\phi^2$ but these terms can
potentially change the action and thus the potential seen by the
radion. In fact from eq.~(\ref{eq:action_Lambda}) it is intuitive that
these corrections should arise since, in a non-expanding background,
$\Lambda=0$, the action should vanish. In particular these additional
contributions will change the boundary conditions on the branes and
could potentially modify the difference in action between the minimum
and the limit $r_0 \to \infty$ which is most important for the
analysis of the phase transition.

Alternatively the system (\ref{eq:phi_eom})-(\ref{eq:Einstein2}) for
$\Lambda = 0$ can be solved by the so-called superpotential
method~\cite{DeWolfe:1999cp,Csaki:2000zn}. Its large advantage is that
it provides {\it exact} solutions since backreactions are
automatically taken into account. Starting from a superpotential
$W(\phi)$ the equations of motion for the choice
\bea
\varepsilon_\alpha\lambda_\alpha (\phi) = W(\phi(r_\alpha)) 
&+&  \frac{\partial W(\phi(r_\alpha))}{\partial \phi} (\phi-\phi(r_\alpha))
+ \varepsilon_\alpha\gamma_\alpha (\phi-\phi(r_\alpha))^2~,\label{eq:super_lam}\\
V(\phi) &=& \frac18 \left(\frac{\partial W}{\partial \phi}\right)^2 
- \frac{\kappa^2}6 W(\phi)^2~, 
\eea
[where $\varepsilon_{1,2}\equiv \pm1$ refer to the two branes, at
$r_1=0$ and $r_2=r_0$ respectively, according to the $\mathbb
Z_2$-orbifold boundary conditions] can be recast in term of the
first-order differential equations
\be
\phi^\prime = \frac12 \frac{\partial W}{\partial \phi}~,\quad
A^\prime = -\frac{\kappa^2}{6} W~.
\ee
In particular the superpotential of the form
\be
\label{eq:W_def}
W = \frac{12M^3}{l} +k_- \phi^2~,
\ee
provides the following scalar potential 
\be
\label{lll}
V = - \frac{12M^3}{l^2} + 
\frac{ (k_- l)^2 - 4k_- l }{2 l^2}\, \phi^2 
- \frac{(k_- l)^2 \kappa^2}{6 l^2} \, \phi^4~.
\ee
For the aim of estimating $\Lambda$ at order $\phi^2$,
eq.~(\ref{lll}) coincides with the potential (\ref{eq:GW_pot})
after imposing eq.~(\ref{eq:kpm}) and requiring $\phi^2 \ll M^3$. The
corresponding solutions to the equations of motion
read~\cite{DeWolfe:1999cp, Csaki:2000zn}
\bea
A &=& -\frac{r}{l} - \frac{1}{6} v_1^2 e^{2 k_- r}~,\\
\label{eq:phi_superV}
\phi &=& v_1 e^{k_- r}~.
\eea
Consequently once one chooses $v_2$ the brane distance is fixed by
\be
\xi_- = \frac{v_1}{v_2} e^{k_- r_-} = 1~,
\ee
which differs from eq.~(\ref{eq:ex_GW}). Furthermore notice that this
solution is based on the requirement $\Lambda=0$, which means that the
action at its extremum $\xi=\xi_-$ vanishes [{\it
cf.}~eq.~(\ref{eq:action_Lambda})]. On the other hand in the limit $r_0
\to \infty$ the $\phi$ profile (\ref{eq:phi_superV}), which matches with
(\ref{eq:phi}), is still a solution of the equations of motion with
$\Lambda=0$ and thus the radion potential also approaches asymptotically a
vanishing cosmological constant. Hence the
superpotential method shows that backreactions can have an important
impact on the radion potential.

\section{Radion potential including backreactions\label{sec:bk-react}}

The superpotential method can also be generalized to non-vanishing
cosmological constant~\cite{DeWolfe:1999cp} and in principle every
solution to the equations (\ref{eq:phi_eom})-(\ref{eq:Einstein2}) can
be exactly derived from some superpotential. Nevertheless it is not
of much use in determining the radion potential since, for fixed
scalar bulk and brane potentials, it cannot be used to find several
solutions corresponding to different brane separations~\footnote{
Using the superpotential method a change in the brane separation would lead,
according to (\ref{eq:super_lam}), to a change in the brane
potentials.}. Understanding the structure of the radion potential
(partially) including backreactions is the aim of this section.

For an arbitrary brane separation the system
(\ref{eq:phi_eom})-(\ref{eq:Einstein}) does not always have a
solution. There are three integration constants, the brane separation
$r_0$ and the parameter $\Lambda$ and four constraints on the branes
(\ref{eq:brane_constraints}). However only the combination $\Lambda
e^{2A}$ enters in the equations and one integration constant can be
eliminated. It can be used to choose \eg $A(0)=0$ or $|\Lambda|=1$. In
summary we have four constants to be fixed by four boundary
conditions, so generically one expects a unique solution for given
bulk and brane potentials. This is not too surprising since one would
expect that the system does not allow for a time-independent solution
(up to Hubble expansion) for the radion field when it is not located
at an extremum of the potential.

Naively one would like to determine the action for several brane
separations and identify it with the negative potential seen by the
radion. This is basically the procedure followed by Goldberger and
Wise in a fixed gravitational background. However there are two
objections to calculating the radion potential in this way if
backreactions are included. First, solutions to the Einstein equations
do not constitute an extremum of the Einstein-Hilbert action, so the
action of the gravitational part should have no physical
significance. This problem can be easily overcome by including the
Gibbons-Hawking term in the action~\cite{York:1972sj, Gibbons:1976ue}
as we will see. Second, the system of equations
(\ref{eq:phi_eom})-(\ref{eq:Einstein2}) only allows for brane
separations that correspond to extrema in the radion potential.  One
way of avoiding this latter problem would be to solve all equations
including a time-dependence and this program is followed in the
vicinity of the static solution in ref.~\cite{Csaki:1999mp}, though
without any stabilization mechanism for the radion.

In the present work we will present an alternative and simplified treatment
which includes the bulk potential. We first determine the action and the radion
mass in the extrema of the radion potential and, after
computing the kinetic term, we use this information to get a
reliable fit to the whole potential. We finally check
that the results of our effective four-dimensional theory are
consistent with the ones of ref.~\cite{Csaki:1999mp}.

We will focus on the particular case of the bulk and brane
potentials (\ref{eq:GW_pot})-(\ref{eq:brane_pot}) constrained by the
bounds (\ref{eq:constraints}) and with solution (\ref{eq:phiApprox}).
However the rationale we follow can be used to analyze the radion
potential also in other scenarios.

\subsection{Cosmological constant and position of the extrema}

The values of the action at its extrema are related to the expansion
parameter $\Lambda$ at those points according to
(\ref{eq:action_Lambda}). We will determine this parameter in this
section. A basic ingredient for our analysis is the scalar solution (\ref{eq:phiApprox})
which was obtained under the
constraints~(\ref{eq:constraints}). Besides we make use of stiff scalar
potentials on the branes. In this case the scalar field is fixed to
the values $v_\alpha$ on the branes and eq.~(\ref{eq:Einstein2}) in
combination with the boundary conditions (\ref{eq:brane_constraints})
reads
\be
\label{eq:LamConstraint}
\left. \Lambda e^{-2 A(r_\alpha)} 
+ \frac{\kappa^2}{12} \phi'^2 \right|_{r=r_\alpha}
=  \frac{\kappa^4}{36} \lambda_\alpha^2 (v_\alpha)
+ \frac{\kappa^2}6 V(v_\alpha)~.
\ee
On the basis of this equation it is possible to accurately determine
the expansion parameter~$\Lambda$ including~\footnote{Subject to 
the constraints in (\ref{eq:constraints}).}
backreactions. Notice that due to the presence of stiff brane potentials only the
left-hand side depends on the brane separation, once the model
parameters are fixed, while the right-hand side can be arbitrarily
chosen due to the brane potentials.  In the following we will first discuss
the case where the brane potential is tuned to reproduce the results
obtained with the superpotential method for the choice
\be
\varepsilon_\alpha\lambda_\alpha^0 =  \frac{12 M^3}{l} + k_- v_\alpha^2~,
\ee
and we will subsequently detune it to obtain cosmologically
more realistic potentials as the one deduced by Goldberger and Wise.

\subsubsection{Tuned case}

\begin{figure}[b]
\psfrag{lll}[][]{\footnotesize$\Lambda\, l^2/10^{-7}$}
\psfrag{rrr}[][]{\footnotesize$r_0/l$}
\includegraphics[width=0.65\textwidth, clip ]{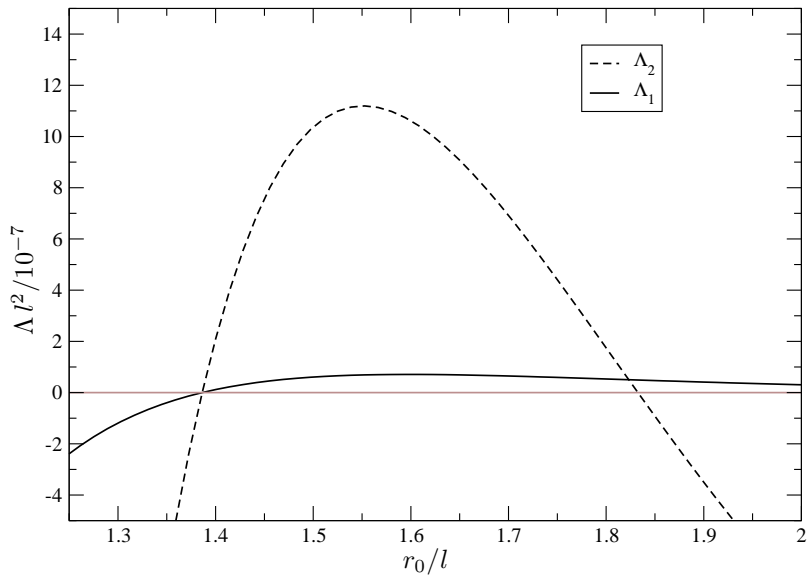}
\caption{\label{fig:Lams} 
The functions $\Lambda_{1,2}$ leading to the two different solutions
of the equations of motion. The parameters used are $k_-l = 0.5, \, v_1=0.1
M^{3/2}, \, v_2 = 0.05 M^{3/2}\,\text{and} \, l= M^{-1}$ which corresponds to $m^2
l^2 = 2.25$. The values are chosen for illustrative purposes and do
not lead to a realistic hierarchy.}
\end{figure}

Let us consider eq.~(\ref{eq:LamConstraint}) evaluated for the two
branes at $r \in \{ 0,r_0 \}$. For the brane potentials
$\lambda_\alpha^0$ used in the superpotential method these two
equations read
\bea
\label{eq:lambs}
24 M^3 \Lambda_1 &=&  v_1^2 k_-^2 - 
(v_1 k_- + k_+ (v_2 - v_1 e^{k_- r_0}) e^{-k_+ r_0}  )^2~,     \nn \\
24 M^3 \Lambda_2 e^{2 r_0}&=&  v_2^2 k_-^2 - 
(v_1 k_- e^{k_- r_0} + k_+ (v_2-v_1 e^{k_- r_0}) )^2~,     
\eea
where we have made use of the equality
\be
m^2=k_-^2-4k_-/l~.
\ee
Owing to the choice of brane potentials two solutions with vanishing
expansion parameter, $\Lambda=0$, are given by $r_0 \to \infty$ and
$r_0=r_-$ with
\be
\frac{v_1}{v_2} e^{k_- r_-} = 1~,
\ee
as obtained by the superpotential method. Nevertheless there is an
additional solution with a larger brane separation and a positive
cosmological constant that leads to a positive Einstein-Hilbert action
according to eq.~(\ref{eq:action_Lambda}).
As an example we show in fig.~\ref{fig:Lams} the functions
$\Lambda_{1,2}$ versus $r_0$ corresponding to a given set of parameters. 

An analytic estimate for the second solution can be obtained as
follows. Notice that the function $\Lambda_1$ is small compared to
$\Lambda_2$ due to the factor $e^{-k_+ r_0}$ and
\be
\label{eq:Lam1}
24 M^3 \Lambda_1 \approx - 2 k_- k_+ v_1 (v_2 - v_1 e^{k_- r_0}) e^{-k_+ r_0}~.
\ee
Hence the values of $r_0$ for the solutions to $\Lambda_1 =
\Lambda_2$ are close to the zeros of $\Lambda_2$ that are given by
\be
\label{eq:min_max_super}
\xi_\mp \equiv \frac{v_1}{v_2} e^{k_- r_\mp} = \frac{k_+ \mp k_-}{k_+ - k_-}~.
\ee
The first solution is just the usual solution obtained by the
superpotential method
\be
r_- = \frac{1}{k_-} \log \frac{v_2}{v_1}~,
\ee
while the other solution is given by
\be
\label{eq:r_plus}
r_+ = r_- + \frac{1}{k_-} \log \frac{k_+ + k_-}{k_+ - k_-}
\approx r_- + \frac2{k_+}\approx r_- + \frac{l}2~.
\ee
Using this value in eq.~(\ref{eq:Lam1}) gives for $\Lambda$ the value
\be
\label{eq:lambda_finetun}
\Lambda_+ \approx \frac{k_-^2}{6} \frac{ v_1 v_2 }{M^3} e^{-4 r_+/l}~.
\ee
The expansion parameter $\Lambda_+$ is in this case additionally
suppressed compared to the energy scale involved in the hierarchy
problem and one might attempt to use this fact to solve, or at least
to alleviate, the cosmological constant problem in a similar way to
the proposals in ref.~\cite{warpedCC}. However and not
unexpectedly, as we will find in a later section, this solution does
correspond to a maximum in the radion potential and is not stable.

Let us compare this result with the findings from the superpotential
method. Since backreactions are taken into account to order $\phi^2$
the action and the brane separation $r_-$ should agree to this order
and they indeed precisely do. Notice that the expansion parameter
$\Lambda$ for an infinitely large brane separation also vanishes in
agreement with the superpotential method. One can thus infer that the
radion potential has two degenerate minima at $r_0 = \{ r_-,\infty \}$
separated by a maximum at $r_+$~\footnote{A more detailed explanation
why this guess is correct will be provided in the next
subsection.}. Consequently a realistic phase transition proceeding
from $\infty$ to $r_-$ requires some detuning away from the scenario
with a superpotential, \eg by changing the brane potentials which is
the next step of our analysis.

\subsubsection{Detuned case}

Using the brane potentials of the tuned case above does not allow a
realistic phase transition. Since the radion potential is degenerate
at $r_0=r_-$ and $r_0\rightarrow \infty$ the system would be stuck in
the high temperature phase. For a realistic model we should modify these two
configurations by some detuning of the UV and IR brane tensions.
Following the same steps leading to eqs.~(\ref{eq:lambs}) the new boundary
conditions for $A^\prime$ in eq.~(\ref{eq:brane_constraints}) lead to
\bea
\label{eq:Lam_detuned1}
24 M^3 \Lambda_1 &=&  v_1^2 k_-^2 - 
(v_1 k_- + k_+ (v_2 - v_1 e^{k_- r_0}) e^{-k_+ r_0}  )^2 +c_1
v_1^2~,      \\
\label{eq:Lam_detuned2}
24 M^3 \Lambda_2 e^{2 r_0}&=& v_2^2 k_-^2 - (v_1 k_- e^{k_- r_0} + k_+
(v_2-v_1 e^{k_- r_0}) )^2+ c_2 v_2^2~, 
\eea 
where we have parameterized the detuning by $c_\alpha v_\alpha^2 \equiv
\frac23 (\lambda_\alpha^2 -{\lambda^0_\alpha}^2)\,$. 

We first enforce
that today's observed expansion is small. If the minimum of the
potential has a vanishing expansion parameter $\Lambda$, according to
eq.~(\ref{eq:Lam_detuned2}) it is given by
\bea
\label{eq:minimum}
\xi_- = \frac{v_1}{v_2} e^{k_- r_-} = \frac{k_+ + \sqrt{k_-^2+c_2}}{k_+ - k_-}~,
 \eea
which depends on the free parameter $c_2$. For $\Lambda=0$
eq.~(\ref{eq:Lam_detuned1}) reads
\bea
\label{eq:Lambda_approx}
0 = 24 M^3 \Lambda_1 &\approx& 2 v_1 v_2 k_+ k_- 
\left( \xi  - 1 \right) e^{-k_+ r}
+c_1 v_1^2 ~,
\eea
which together with the position of the minimum $\xi_-$ in
eq.~(\ref{eq:minimum}) fixes the value of $c_1$ as
\be
\label{eq:c1}
c_1 v_1^2 = 2 v_1 v_2 k_+ | k_- | ( \xi_- - 1) \, e^{-k_+ r_-}~.
\ee
This choice is the fine-tuning that is needed to solve the
cosmological constant problem and it is in general present in RS-type
models.

Once one fixes $\xi_-$ (and correspondingly $c_1$ and $c_2$) the system
(\ref{eq:Lam_detuned1})-(\ref{eq:Lam_detuned2}) has a second solution
$\xi_+$ that fulfills
\bea
\label{eq:maximum}
&&k_-^2 + c_2 - (k_- \xi_+ + k_+ (1-\xi_+))^2 \approx
c_1 \frac{v_1^2}{v_2^2} e^{2 r_+/l}  \nn \\
&&= 2 \frac{v_1}{ v_2} k_+ | k_- | ( \xi_- - 1) \, e^{-k_+ r_- + 2 r_+/l}~.
\eea
As long as $c_2$ is small the right-hand side can be neglected and the
position of $\xi_+$ is, in a similar way to the minimum $\xi_-$, given
by
\bea
\label{eq:maximum2}
\xi_+ = \frac{v_1}{v_2} e^{k_- r_+} = 
\frac{k_+ - \sqrt{k_-^2+c_2}}{k_+ - k_-} =
\frac{2 k_+ }{k_+ - k_-} - \xi_- ~. 
\eea
On the other hand if $c_2$ surpasses $k_+^2 - k_-^2$ the left-hand
side is positive for $\xi_+>0$. In this case, the right-hand side
has to become comparatively large what happens at $r_+ \approx
\frac{k_+ l}{2}r_- \approx 2 r_-$ which implies $\xi_+ \approx
\frac{v_2}{v_1} \xi_-^2 < \xi_-$ for $k_-<0$. Comparison shows
that sizeable deviations from (\ref{eq:maximum2}) are hence not
expected for
\be
\label{eq:bound_xim}
\xi_- \lesssim - \frac{v_1}{2 v_2} + \sqrt{\frac{ v_1^2}{4
v_2^2} + \frac{2 v_1}{ v_2}} ~.
\ee
It turns out that the parameter space that violates this bound is less
interesting in the sense that it leads generally to rather strong
phase transitions. In this way we will mostly be concerned with the
region in parameter space where (\ref{eq:maximum2}) is a good
approximation.

Plugging eq.~(\ref{eq:c1}) in (\ref{eq:Lam_detuned1}) one finds the
expansion parameter at the maximum $\xi_+$ to be
\be
\label{eq:lambda_+}
 \Lambda_+ \approx  \frac{v_1 v_2}{12 M^3} k_+ | k_- | 
\left( ( \xi_- - 1 ) e^{-k_+ r_-} + ( 1 - \xi_+ ) e^{-k_+ r_+}
\right)~.
\ee
If $\xi_-$ is not too close to unity the first contribution dominates
this expression because $r_+ > r_-$. On the other hand the
cosmological constant in the limit $\xi \to 0$ ($r_0 \to \infty$) is
given by this first contribution
\be
\label{eq:lambda_inf}
 \Lambda_\infty \approx  \frac{v_1 v_2}{12 M^3} k_+ | k_- | 
( \xi_- - 1 ) e^{-k_+ r_-}~.
\ee
as it can be easily deduced from eqs.~(\ref{eq:Lam_detuned1}) and
(\ref{eq:c1}). The radion potential as a function of $\xi$ is hence
very flat between $\xi=0$ and $\xi=\xi_+$ and then quickly drops to
zero at $\xi=\xi_-$. In a subsequent subsection we will provide a
parametrization that interpolates between these extremal values.

We will also demand that this construction solves the hierarchy
problem, \ie $r_- \approx 37\, l$, and thus for fixed ratio $v_1/v_2$
the parameter $m^2$ (and hence $k_-$) should be accordingly chosen
while we treat $\xi_-$ (or $r_-$) as a free parameter. In principle
$c_2$ can always be chosen such that one obtains for $\xi_-$ an
arbitrary value in the interval $[1,\infty]$ (with $\Lambda_\infty\ge0$)
but we will assume $\xi_-$ to be smaller than $v_1/v_2$ in order to
obtain a negative $k_-$
\be
k_- = \frac{1}{r_-} \log\left( \frac{v_2}{v_1} \xi_-\right)~.
\ee

About the approximations we employed, notice that we used
(\ref{eq:GW_pot}) as scalar potential and nowhere the assumption
$\phi^2 \ll M^3$ that needs to be fulfilled in order to make contact
with the superpotential method. The sole constraint for the
applicability of our results so far is (\ref{eq:constraints}). This
leads to the inequalities [using (\ref{eq:phiApprox})]
\be
\label{eq:vel_bound2}
m^2 l^2 \phi^2 \lesssim m^2 l^2 v_1^2 \approx - 4 l k_- v_1^2  \ll 24 M^3~,
\ee
and 
\be
\label{eq:vel_bound}
l^2 \phi^{\prime 2} = v_2^2 \lp l k_- \xi + l k_+ (1-\xi) \rp^2
< v_2^2 \lp l k_- \xi_- + l k_+ (1-\xi_-) \rp^2 \ll 24M^3~.
\ee
Depending on the parameters these constraints can be much weaker than
those employed in the literature to ensure smallness of
backreactions~\cite{Goldberger:1999uk, Creminelli:2001th} namely
$\phi^2 \ll M^3$. However, they still guarantee that the
deformation of the CFT induced by the operators corresponding to the
bulk scalar can be treated perturbatively at all scales up to the 4D
Planck scale~\footnote{In the case $k_- >0$ these operators become
  strong in the IR which is potentially more
  cumbersome~\cite{Creminelli:2001th}.}.

Notice that even if the conditions (\ref{eq:vel_bound2}) and
(\ref{eq:vel_bound}) are fulfilled (which ensures that the relative
error in $A'$ is small) there still can be sizeable (cumulative)
deviations in $e^A$ since $A(r_0) \approx 37$. Nevertheless this will
not affect the equations of motion of $\phi$ that only depend on $A'$
and not on the exponential warp factor. Hence the expansion parameter
deduced from the equations
(\ref{eq:Lam_detuned1})-(\ref{eq:Lam_detuned2}) is reliable. Still the
detuning parameters need in principle (small) corrections to reproduce
the correct hierarchy. In the following we call the regime where our
constraints are saturated and our approximation becomes unreliable the
regime of {\it large backreactions}.

Let us compare these findings with the results from the superpotential
method and those in the GW framework. The GW potential can be
written as
\be
V_{GW} = l^{-1} 
v_2^2 \lp \frac{v_1}{v_2} \xi \rp^{-4/(l k_-)}
\left[ (4- lk_-) (1 -\xi)^2 + lk_- \right] + \textrm{const}~,
\ee
and the extrema fulfill
\be
\frac{4}{lk_- \xi_\pm} = \frac{2(4- lk_-) (1-\xi_\pm)}
{(4- lk_-) (1 -\xi_\pm)^2 + lk_- }~.
\ee
This leads to a quadratic equation for $\xi_\pm$ with solutions that
are of the form
\be
\xi_\pm = \frac{k_+ \mp \sqrt{k_-^2 + c_2}}{k_+ - k_-}~,
\ee
for an appropriate $c_2$. This coincides with the positions of the
extrema in the detuned case found in (\ref{eq:minimum}) and
(\ref{eq:maximum2}) under the assumption
(\ref{eq:bound_xim}). The value of the potential in the extrema is
given by
\be
\label{eq:pot_diff}
V_{GW}(r_\pm) - V_{GW}(\infty) = 2 v_2^2 l^{-1} 
e^{-4 r_\pm
/l} (l k_-) \xi_\pm (1-\xi_\pm)~,
\ee
which is proportional to the difference in expansion parameters
$(\Lambda_\pm-\Lambda_\infty)$. The value of the radion potential can
indeed be inferred from the expansion parameter $\Lambda$ as we will
see in section~\ref{sec:kin_pot}.  In conclusion our results agree [at
least to order $\mathcal O(\phi^2)$] with the superpotential method
for the choice
\be
\xi_-=1 \quad (\textrm{superpotential method})\,,
\ee
while the GW potential (up to a constant piece) is
reproduced for the choice 
\be
\label{eq:xi_GW}
\xi_- \simeq 1 + |l k_-|^{1/2}/2 - l k_-/4
\quad (\textrm{Goldberger-Wise})\,,
\ee
under the assumption (\ref{eq:bound_xim}). Hence our approach
unifies the results obtained both in the GW and superpotential
approaches. 

Finally some comments on the role of the original GW
potential are in order. In our framework the potential obtained by
Goldberger and Wise merely corresponds to a specific choice for the
parameter $\xi_-$. From our point of view there is no special
significance to this choice and it just results from imposing
initially the same tension on the two branes when the backreactions
are ignored. Once the behavior of the scalar field is determined one
would be forced to adjust these brane tensions in order to cancel the
contributions from the backreactions and to obtain a vanishing
expansion parameter at the minimum (as already mentioned in
ref.~\cite{Goldberger:1999uk}). Hence this parameter choice is in no
way a distinguished one. In particular the fact that the potential
difference (\ref{eq:pot_min_GW}) scales with $(l k_-)^{3/2}$ results
from the peculiar choice in (\ref{eq:xi_GW}) (or equivalently for the
brane potentials). This scaling is also reflected in the radion mass
as we next discuss.


\subsection{The radion mass \label{sec:rad_mass}}
In this subsection we will determine the radion mass in the detuned case
along the lines of ref.~\cite{Csaki:2000zn} while the decoupling of the
linearized Einstein equations is demonstrated in appendix~\ref{sec:appx}. It
turns out that in the case of an expanding Universe the impact of the
expansion parameter $\Lambda$ can be absorbed in the radion mass parameter
\be
\hat m_{rad}^2 = m_{rad}^2 + 6 \Lambda~.
\ee
However the contribution from $\Lambda$ is anyway negligible (being
smaller by a factor $e^{2A}$) leading to the equation [see
eq.~(\ref{eq:app_final})]
\be
\hat F'' + 2 A'' \hat F - 2 A' \hat F' - 
2 \frac{\phi_0''}{\phi_0'} \hat F' = - m_{rad}^2 \, e^{-2A} \hat F ~,
\ee
which can be solved to obtain $m_{rad}$.  If backreaction is
neglected, $A''=0$, this allows for a solution with $\hat F=1$ and
$m_{rad}^2=0$. For small backreaction, $\hat F= 1 + f$, this system of
equations can be linearized as
\bea
\label{eq:Flin}
f'' - 2 A' f' - 2 \frac{\phi_0''}{\phi_0'} f' 
&=& -2 A'' -  m_{rad}^2 \, e^{-2A} \nn \\
&=& \frac{2 \kappa^2}3 \phi_0'^2 - m_{rad}^2 \, e^{-2A} ~,
\eea
while the boundary conditions in the limit $\gamma_\alpha\to\infty$
read $f'(0)=f'(r_0)=0$. The main difference with respect to the
standard solution presented in ref.~\cite{Csaki:1999mp} comes from the
factor $\phi_0''/\phi_0'$ that deviates from $k_-$ close to the TeV
brane.

The solution reads
\be
\label{eq:f_sol}
f'(r) = \phi^{\prime2}(r) \, e^{2 A(r)}
\int_0^r dx \, \lp
\frac{2\kappa^2}3 e^{-2 A(x)} 
- \frac{m_{rad}^2}{\phi^{\prime2}(x)} e^{-4A}\rp~,
\ee
that by construction fulfills the boundary condition $f'(0)=0$. At
this point we will neglect backreactions in $A$ that would lead to
corrections of order $O(\phi^4)$ on the radion mass~\footnote{When
approaching the region of large backreactions one would have to take
the above mentioned cumulative effect in the warp factor into account
and the final radion mass would involve the corrected warp factor
instead of the plain one. Notice that the determination of the radion
mass only serves as an additional check on our radion effective action
and the tunneling analysis of the radion is generally not affected by
this subtlety.}. Then the constraint $f'(r_0)=0$ determines the mass
by the equation
\be
m_{rad}^2 \int_0^{r_0} dx \, \frac{e^{4 x/l}}{\phi^{\prime\,2}(x)} 
= l \frac{\kappa^2}3 e^{2r_0/l}~.
\ee
For the scalar field solution in the limit of small
backreactions (\ref{eq:constraints}), which can be written
\be
\phi'(r) = v_1 k_- e^{k_- r} 
\left( 1 - \tilde q \, e^{(k_+ - k_-) (r - r_0)} \right ),
\quad \tilde q= k_+/k_- \left( 1- \frac1{\xi} \right)~,
\ee
the above integral becomes (using $l k_+ = 4 - l k_-$)
\be
\int_0^{r_0} dx \, \frac{e^{4 x/l}}{\phi^{\prime\,2}(x)} 
\approx \frac{l\,e^{(4 - 2 lk_-) r_0/l}}{4 v_1^2 k_-^2 (1- \tilde q)}~. 
\ee
For $\tilde q > 1$ the integrand actually contains a pole. However a
more sophisticated analysis shows that the solution (\ref{eq:f_sol})
is still regular due to the pre-factor $\phi^{\prime2}(r)$ and that
the naive integration is justified. Therefore one concludes that the
radion mass is
\bea
\label{eq:m_approx}
m_{rad}^2 &=& \frac23 k_-^2  
 (1 - \tilde q ) e^{(-2 + 2lk_-) r_0/l} \frac{v_1^2}{M^3}\nn \\ 
&\approx&  \frac8{3l^2}  l k_-  
\lp 1 - \xi + \xi \frac{k_-}{k_+} \rp \xi\, e^{-2 r_0/l} \, \frac{v^2_2}{M^3}~.
\eea

Observe that in the case without detuning (corresponding to a scenario
that can be treated with the superpotential method)
\be
\xi_\pm = \frac{k_+ \pm k_-}{k_+ - k_-}~,
\ee
the masses turn out to be
\be
\label{eq:m_approx_tuned}
m_{rad,\pm}^2 \approx \mp \frac{2}{3l^2} (lk_-)^2 
\, e^{-2 r_\pm/l} \, \frac{v^2_2}{M^3}
\quad (\textrm{superpotential})\,,
\ee
which agrees with the physical radion mass $m_{rad}^2$ found in
\cite{Csaki:2000zn} and also with the interpretation that $\xi_-$
denotes a minimum of the potential while $\xi_+$ is a
maximum. Moreover in the case of GW, see eq.~(\ref{eq:xi_GW}), the
mass is
\be
m_{rad}^2 \approx \frac4{3 l^2} |l k_-|^{3/2}  
\, e^{-2 r_0/l} \, \frac{v^2_2}{M^3}
\quad (\textrm{Goldberger-Wise})\,,
\ee
which indeed scales with $|lk_-|^{3/2}$ as expected from the GW
potential in eq.~(\ref{eq:GW_pot}).

Notice that the radion mass is slightly below the TeV scale what is
essential for an effective four-dimensional description. To compare
the radion mass obtained here with the one from the radion potential
involves the kinetic term which is the topic of the next subsection.


\subsection{Kinetic term \label{sec:kin_pot}}

So far we have only discussed the occurring expansion parameter
$\Lambda$ for the different solutions of the equations of motion. The
main motivation was the relation (\ref{eq:action_Lambda}) that implies
that the action is proportional to the expansion parameter. In the
current subsection we will make contact between the five-dimensional
system and the effective action of the radion.

Let us start with the kinetic term of the radion. It was derived in
several ways~\cite{Goldberger:1999un, Csaki:1999mp, Rattazzi:2000hs}
and here we briefly review the calculation of
ref.~\cite{Csaki:2000zn}. We will use the metric (\ref{eq:metric_pert}) but
will neglect the contributions from the scalar field. These effects are
suppressed by a factor $(l k_-)^2 v_1^2/M^3$~\cite{Csaki:2000zn} which
is small under the constraints~(\ref{eq:vel_bound}). The Einstein
equations are then solved by the radion Ansatz $F(x,r) = e^{2r/l}
R(x)$ and the Einstein-Hilbert action contains a kinetic term of the
form
\be
K_{rad} = 6 M^3 l \int d^4x 
\sqrt{|\det \bar g_{\mu\nu}|}\left( e^{2 r_0/l} - 1 \right) (\partial R)^2~.
\ee
The correct normalization is obtained by the observation that in this
background the geodesic distance of the branes is given by
\be
r(x) = \int_0^{r_0} dr \lp 1 - 2e^{2r/l} R(x) \rp
= r_0 - l \, e^{2 r_0/l}  R(x)~,
\ee
from which $(\partial R)^2 = l^{-2} e^{-4 r_0/l}(\partial r)^2 \simeq
l^{-2} e^{-4 r/l}(\partial r)^2$.  Hence $K_{rad}$ becomes
\bea
K_{rad} &=& 12 M^3 l^{-1} \int d^4x \sqrt{| \det \bar g_{\mu\nu}|} 
 \;\frac12
 (\partial r)^2
 e^{-2 r / l}  \nn \\
&=& 12 (M l)^3 \int d^4x \sqrt{| \det \bar g_{\mu\nu}|} 
\; \frac12
 \left(\partial \mu\right)^2
,
\eea
with the notation $\mu = l^{-1} e^{-r/l}$. We do not absorb the
pre-factor $12(ML)^3$ in the definition of $\mu$ because it will prove to be
useful in the analysis of the phase transition~\footnote{Notice that
this kinetic term was obtained by decoupling the Einstein equations
and the result is a factor of 2 smaller than the one from the more
naive approach derived in~\cite{Goldberger:1999uk, Creminelli:2001th}
and subsequently used in the analysis of the phase transition
in~\cite{Creminelli:2001th, Randall:2006py, Nardini:2007me}.}.

In ref.~\cite{Goldberger:1999uk} the action resulting from the bulk
scalar was right away interpreted as the negative potential seen by
the radion. In the present case this could hardly be true since the
action is positive for an expanding Universe. We better expect this to
correspond to a maximum and not to a minimum of the potential, since
on the other hand the radion mass is negative in this
situation. Moreover an expanding Universe without any bulk scalar
would also have a non-vanishing action, even though the radion does
not see any potential in this case.

In order to obtain the potential seen by the radion we must separate
the action into the contributions from the expansion and that from the
radion interacting with the bulk scalar. The contribution to the
action from the expansion is according to (\ref{eq:R_scalar}) given by
\be
S_{exp} = 12 M^3 \int d^5x \sqrt{| \det g_{MN} |} 
e^{-2A} \Lambda~,
\ee
and, considering the action provided by (\ref{eq:action_Lambda}), it
results in a radion action
\be
\label{eq:remainder}
S_{rad} = -6 M^3 \int d^5x \sqrt{| \det g_{MN} |} 
e^{-2A} \Lambda~,
\ee
which must arise from the value of the radion potential at its extrema.

As mentioned before an additional problem arises from the fact that
the solution is not an extremum of the action in general relativity.
To overcome this problem one can confine the system to a box, $t \in
[-T,T]$, and add the so-called Gibbons-Hawking~\cite{Gibbons:1976ue}
term to the action~\footnote{ Inclusion of this term is actually not
relevant since the radion potential could be inferred from
(\ref{eq:remainder}), but it will elucidate the comparison with
ref.~\cite{Csaki:1999mp} later on.}
\be
S_{GH} = \int_{\partial M} 2 K~,
\ee
where $K$ denotes the extrinsic curvature and $\partial M$ is the
boundary of the space-time manifold. Using the metric given in
(\ref{eq:def_g}) one can evaluate the Gibbons-Hawking
term~\cite{Wald:GR}
\be
\label{eq:Gibbons_Hawking}
S_{GH} = - 18 M^3 \int d^5x \sqrt{| \det g_{MN} |} 
 e^{-2A} \Lambda~,
\ee
and the total action is given by
\be
\label{eq:action_tot}
S_{tot} = S_{exp} + S_{rad} + S_{GH} = - 12 M^3 \int d^5x \sqrt{| \det g_{MN} |} 
e^{-2A} \Lambda~. 
\ee
After integration over the fifth dimension the effective action for
the radion hence reads~\footnote{Note that we integrate over the
orbifold $S_1/\mathbb{Z}_2$ and hence twice over the bulk as it is
customary in the literature on RS models.}
\bea
\label{eq:action_eff}
S_{eff} &=& 12 (M l)^3 \int d^4x \sqrt{| \det \bar g_{\mu\nu}|} 
\left( \frac12
 (\partial \mu)^2  - V_{rad}(\mu)
\right)  \nn \\
&& + M_P^2 \int d^4x \sqrt{| \det \bar g_{\mu\nu}|} R_{(4)} + S_{GH}~,
\eea
where we defined the four-dimensional Planck mass, $(M_P l)^2 = (M
l)^3$, and identified the radion potential in its extrema as
\be
\label{eq:pot_cosm}
V_{rad}(r_\pm) =l^{-2} \Lambda_\pm/2~.
\ee
This is the essential relation that we use to connect the value of the
radion potential to the expansion around its extrema.

Let us compare this result with the one obtained in
ref.~\cite{Csaki:1999mp}. There an effective action was obtained by
perturbing the background metric but without taking a stabilizing
mechanism into account. Their result reads in our notation ($a$ is the
scale factor)
\bea
S_{eff} &\propto& \int dt \, a^3 
\left( \frac12 \dot r^2 e^{-2r/l} l^{-2} 
+ \frac12 \lp\frac{\ddot a}{a} + \lp \frac{\dot a}{a}\rp^2 \rp 
- V_{rad}(r)\right) + S_{GH} \nn \\
&=& \int dt \, a^3 
\left( \frac12  \dot r^2 e^{-2r/l} l^{-2} 
-  \frac12  \lp \frac{\dot a}{a}\rp^2  - V_{rad}(r) \right) ~.
\eea
The last equality is obtained by partial integration which cancels the
Gibbons-Hawking term at a time-like boundary as discussed above. The
radion potential $V_{rad} (r)$ was not specified in~\cite{Csaki:1999mp}
but comparison with our result (\ref{eq:action_eff}) confirms that the
radion potential and the expansion (including the Gibbons-Hawking
term) contribute equally to the action in the extrema of the radion
potential.

\subsection{Interpolating potential}

In the following we will approximate the effective potential seen by the
radion. The equations of motion have only stationary (up to the Hubble
expansion) solutions in the three extremal situations $r \in\{ r_-,
r_+, \infty\}$ [or $\xi \in\{ \xi_-,
\xi_+, 0\}$] and we use these three values to provide a fit to the
potential. The positions of the extrema of the radion potential are
given by eqs.~(\ref{eq:minimum}) and (\ref{eq:maximum}).  We
parametrize the potential by
\be
V_{rad}^\prime(\xi) \propto  \xi^{\omega-1} (\xi-\xi_-) (\xi- \xi_+)~,
\ee
and integration yields
\be
\label{eq:fitV}
V_{rad}(\xi) \propto  \xi^\omega \lp \frac{\xi^2}{\omega+2}  -  
\frac{\xi (\xi_-+ \xi_+)}{\omega+1}  + \frac{\xi_- \xi_+}
{\omega}\rp~,
\ee
where the integration constant is fixed to zero by $V_{rad}(0)=0$.
The parameters $\omega$ and the pre-factor can be adjusted in order to
reproduce $V_{rad}(\xi_\pm)$ given by eqs.~(\ref{eq:lambda_+}),
(\ref{eq:lambda_inf}) and (\ref{eq:pot_cosm}).  This corresponds to
the expression
\be
V_{rad}(\xi) l^4 =  \lambda \, e^{-k_+ r} P(\xi)~,
\ee
with
\be
 \lambda = \frac1{24} \frac{v_1 v_2}{M^3} l k_- l k_+~,
\ee
and
\be
 P =  \frac{1-\xi_-}{\xi_-}\lp \frac{\xi}{\xi_-} \rp^{\omega+k_+/k_-} 
\frac{\omega(\omega+1) \xi^2
- \omega(\omega+2) (\xi_- + \xi_+)\xi
+ (\omega+1)(\omega+2) \xi_- \xi_+ }
{ \omega(\xi_- - \xi_+) - 2 \xi_+}~.
\ee
The function $P(\xi)$ is normalized to $P(\xi_-)= (\xi_- -1)$ and $\omega$ is
determined by $P(\xi_+)= (\xi_+ - 1)$, which yields
\be
\label{eq:def_omega}
\lp \frac{\xi_+}{\xi_-} \rp^{\omega+k_+/k_-+1}
\frac{\omega(\xi_+ - \xi_-) - 2 \xi_- }
{\omega(\xi_- - \xi_+) - 2 \xi_+ }
= \frac{1-\xi_+}{1-\xi_-}~,
\ee
and hence $\omega \approx -k_+/k_-$.

Another ingredient useful for the fit is the radion mass at the
extrema where the second derivative of $V_{rad}$ turns out to be
\be
\xi^2 \frac{d^2}{d\xi^2} V_{rad} (\xi_\pm) =
\lambda \, \omega^2 \, e^{-k_+ r}\, 
\frac{(1-\xi_\pm)(\xi_- - \xi_+)}{(\xi_- - \xi_+) \pm 2\xi_\pm/\omega}~.
\ee
Using the standard kinetic term for the $\mu$ field
\be
\mu = l^{-1} \, e^{-r/l}~, ~\quad \partial
\xi/\partial \mu = -l k_- \xi/\mu~,
\ee
one gets the expression for the mass
\bea
m^2 &=& V_{rad}^{\prime\prime}(\xi_\pm) 
\left( \frac{\partial \xi}{\partial \mu}\right)^2 \nn \\
&\approx& 16 \,l^{-2}\, \lambda \, e^{(k_- l - 2) r/l} 
\frac{(\xi_\pm-1)(\xi_- - \xi_+)}{(\xi_- - \xi_+) \pm 2\xi_\pm/\omega}~. 
\eea
If $\xi_-$ is not too close to unity the last factor becomes
$(\xi_\pm-1)$ in agreement with (\ref{eq:m_approx}). On the other
hand in the superpotential limit (\ref{eq:min_max_super})
\be
\xi_-\to1, \quad \xi_+ \to 1-2/\omega~,
\ee
the last factor behaves as [using (\ref{eq:def_omega})]
\be
\frac{(\xi_\pm-1)(\xi_- - \xi_+)}{(\xi_- - \xi_+) \pm 2\xi_\pm/\omega} 
\to \mp \frac{1}{ \omega}~,
\ee
in agreement with eq.~(\ref{eq:m_approx_tuned}). This shows that our
fit is indeed consistent with the effective action
(\ref{eq:action_eff}).

The above parameterization is reasonable as long as $\xi_+ < 1$ which
is the case we will consider hereafter. For the $\xi_+>1$ case on the
one hand eqs.~(\ref{eq:lambda_+}) and (\ref{eq:lambda_inf}) imply
$\Lambda_+<\Lambda_\infty$, so that $V_{rad}(\xi_+)$ becomes negative
and at the same time transforms into a minimum according to
(\ref{eq:m_approx}). In this case our Ansatz for the metric
(\ref{eq:def_g}) does not allow for additional solutions between the
two minima at $\xi_\pm$ while according to (\ref{eq:m_approx}) any
additional solution should be a minimum. Therefore our Ansatz in
(\ref{eq:def_g}) produces in this case two local minima of the radion
potential in configuration space but no local maximum. This
complicates the question of what the radion potential might look like
in the case $\xi_+>1$ so that we will disregard this case in the following.

\section{Holographic phase transition\label{sec:pt}}

In this section we will present the discussion of the holographic
phase transition at finite temperature along the lines of
refs.~\cite{Creminelli:2001th, Randall:2006py, Nardini:2007me}~\footnote{A
precursor at zero temperature can be found in~\cite{Cline:2000xn}.}.
At finite temperature the system allows for an additional
gravitational solution with a black hole singularity in the bulk.
This AdS-S metric describes in the AdS/CFT correspondence the high
temperature phase of the system~\cite{Hawking:1982dh, Witten:1998zw}.
This phase starts dominating at temperatures of the order of the TeV
scale. In fact the potential difference between AdS-S and pure AdS
phases is given by~\cite{Creminelli:2001th}
\be
\label{eq:temp_pot}
\left. -4 \pi^4 (M l)^3 T_h^4 \right|_{T_h=T}~,
\ee
where $T_h$ is the scalar field parameterizing the distance between
the horizon and UV brane and $T$ is the temperature of the system. On
the other hand the difference between the RS-GW and pure AdS phases is
expressed by eq.~(\ref{eq:pot_diff}) which should equal
eq.~(\ref{eq:temp_pot}) at the critical temperature $T_c$ of the phase
transition. It turns out that typically $T_c$ is between the
electroweak and TeV scales~\footnote{In reporting the expression
  (\ref{eq:temp_pot}) we omit the subdominant contribution of the bulk
  matter fields since this correction plays a minimal role in our
  analysis. The same consideration holds for the thermal corrections
  to the RS-GW potential. We assume nearly all SM fields to be
  fundamental so that their thermal contribution to the two-phases
  free energies is similar and thus it cancels out.}.

Starting from a hot universe in the AdS-S phase a first-order phase
transition towards RS-GW may happen below $T_c$\,. In the
five-dimensional picture this means that the IR brane emerges from the
black hole horizon and to minimize the bounce action this crossing has
to happen far from the UV brane. For this reason it is commonly
assumed that the bounce path consists in moving the horizon away from
the UV brane till arriving to the (unstable) pure AdS phase, and
subsequently displacing the IR brane from $r=\infty$ to $r=r_-$. This
reasonable assumption fixes the bounce path and it reduces the study
of the tunneling probability to the usual analysis of the
one-dimensional bounce~\cite{Coleman:1977py,Linde:1980tt} in which the
bouncing scalar field is identified with $\mu$ ($T_h$) in the part of
the path between pure AdS and RS-GW (AdS-S)~\footnote{For more details
  about the AdS-S solution and the bounce of this phase transition see
  refs.~\cite{Creminelli:2001th,Kaplan:2006yi,Randall:2006py,Nardini:2007me}.}.

In particular the path in the AdS-S space can be simplified by
observing that the kinetic term of the field $T_h$ has a small
pre-factor~\footnote{The estimate of this factor is controversial. We
checked in the numerical evaluation that admitting a sizeable kinetic
term would only lead to slightly smaller tunneling
temperatures.}. Consequently this part of the path becomes extremely
short once $T_h$ is canonically normalized, so that in the AdS-S
region the potential seen by the bouncing field can be approximated by
a step function.

We will release the radion field $\mu$ from a certain initial
position $\mu_0$ and we will evolve it to the point $\mu=0$ (corresponding
to the pure AdS phase) according to the $O(3)$ bounce
equation~\cite{Coleman:1977py, Linde:1980tt}
\be
\partial_\rho^2 \mu + 2 \frac{\partial_\rho \mu}{\rho} = \partial_\rho
V_{rad}~,
\label{eq:O3}
\ee
where $\rho^2=\vec x^2$.
In the bounce solution the radion field should arrive at $\mu=0$ with
the kinetic energy necessary to jump and stop on the top of the AdS-S
minimum
\be
\left. 4 \pi^4 T^4 = 6 (\partial_\rho \mu)^2 \right|_{\mu=0}~,
\label{eq:jump}
\ee
and this solution is used to determine the bounce action $S_3/T$.

In order to explicitly calculate the tunneling probability it is
useful to rewrite the radion action (\ref{eq:action_eff}) as
\be
S_{eff} = 12\, (Ml)^3 \int d^4x \sqrt{| \det \bar g_{MN}|} 
\left(
\frac12 \dot \mu^2  - V_{rad}(\mu)
\right)~,
\ee
where the potential, neglecting a constant piece, is 
\bea
 V_{rad}(\mu) &=& \hat\lambda \, \mu^4 \, 
\frac{\omega (\omega+1)(\omega+2)}
{ \omega \lp 1 - \frac{\xi_+}{\xi_-} \rp - 2 \frac{\xi_+}{\xi_-}  } 
\lp \frac{\mu}{\mu_-} \rp^{- \omega l k_- -  l k_+ - l k_-} \nn \\
&& \times \quad \left[ \, \frac1{\omega+2} \lp\frac{\mu}{\mu_-}\rp^{-2k_-}
\hskip -0.5cm - \frac1{\omega+1} \lp\frac{\mu}{\mu_-}\rp^{-k_-} \lp 1 + \frac{\xi_+}{\xi_-} \rp 
 + \frac1{\omega}   \frac{\xi_+}{\xi_-}  \right]~,
\eea
with $\omega$ determined by eq.~(\ref{eq:def_omega}) and
\be
 \mu_- = l^{-1}\lp \frac{v_2}{v_1} \xi_- \rp^{- 1/(l k_-)}~,
\ee
\be
\hat\lambda = \frac1{24} \frac{v^2_2}{M^3} (l k_-) (l k_+) (\xi_- -1) \xi_-~.
\ee

The dimensionless tunneling action will only depend on the parameters
$\hat\lambda$, $\omega$ and $\mu_+/\mu_-$ and not explicitly on
$\mu_-$. This can be seen by using the conformal transformation
\be
x^\mu \to a^{-1}x^\mu, \quad
V_{rad}(\mu) \to a^{-4} V_{rad}(a \mu)~, 
\ee
which is equivalent to a rescaling of all dimensionful  quantities, in
particular $\mu_- \to \mu_-/a$. Therefore the relevant scale involved
in the bounce solution is $\mu_0$ and the functional determinant
\cite{Coleman:1977py} of the tunneling process has to be proportional to
$\mu_0^{4}$. Tunneling can hence occur for bubble action $S_b$ 
\be
\label{eq:S3_est}
S_b \simeq \log \frac{\mu_0^4}{\Lambda^2} 
\approx 4 \frac{r_-}{l} + 4 \log \frac{\mu_0}{\mu_-} \lesssim 140~,
\ee
where it is used that the expansion parameter is the cosmological
constant of the RS-GW phase (\ref{eq:lambda_inf}).  Observe that if
the tunneling takes place during radiation domination the Hubble expansion is
quadratic in temperature and the right-hand side of this equation
would increase when the tunneling temperature decreases. However in the
present case the Hubble parameter is dominated by the vacuum energy
(which can even lead to a short period of
inflation~\cite{Linde:1980tt,Nardini:2007me}) and a smaller
temperature in fact lowers the right-hand side via its dependence in
the functional determinant. Notice that even though the
four-dimensional gravity contributions to the action are significant
(due to the expansion) gravitational effects~\footnote{
Gravitational effects become important when nucleated bubbles are
of order $1/\sqrt{\Lambda}$.} in the tunneling process should not
relevantly modify our results~\cite{Coleman:1980aw}.

As long as the release point is far away from the minimum and maximum
of the potential
\be
(\xi_+/\xi_-)^{1/|k_-|} \ll \frac{\mu_0}{\mu_-} \ll 1~,
\ee
the field experiences a nearly conformal potential. If $\xi_-$ is not
too close to unity the position of the maximum, given
by
\be
(\xi_+/\xi_-)^{-1/lk_-} \lesssim \lp \frac{v_2}{v_2} \xi_-
\rp^{-1/lk_-} = e^{-r_-/l} = l \mu_- = 
\frac{\textrm{TeV  scale}}{\textrm{Planck scale}}~,
\ee
is such that there is a large hierarchy between the position of the
minimum and the maximum and for a large range of release points
the potential is nearly conformal.

In the conformal case the potential is of the form
\be
\label{eq:pot_conf}
V_{conf} \to - \bar \kappa \mu^4 ~.
\ee
Solving eqs.~(\ref{eq:O3}) and (\ref{eq:jump}) with this potential
gives an action, temperature and bubble size for the $O(3)$
symmetric bubble
\be
S_3/T \simeq 217.0 \, \bar \kappa^{-3/4} \, (Ml)^3, \quad
T/\mu_0 \simeq 0.103 \, \bar \kappa^{1/4}, \quad
\bar \rho \, \mu_0 \simeq 3.45 \, \bar \kappa^{-1/2}~. 
\ee
However the potential~(\ref{eq:pot_conf}) is normalized to the origin
instead of being normalized at the AdS-S
minimum~\cite{Coleman:1977py} and we have to add to the action
the omitted contribution
\be
\frac{16 \pi^5}{3} T^3 \bar \rho^3 \, (M l)^3 \simeq 72.3 \, 
\bar \kappa^{-3/4} \, (M l)^3~.
\ee
This yields for the tunnel action
\be
\label{eq:tunnel_num}
S_3/ T \simeq 289.3 \, \bar \kappa^{-3/4} \, (Ml)^3~.
\ee

Besides thermal fluctuations the potential barrier can also be
overcome by quantum fluctuations. In this case the bounce solution is
$O(4)$ symmetric and eq.~(\ref{eq:O3}) has to be replaced by
\be
\partial_\rho^2 \mu + 3 \frac{\partial_\rho \mu}{\rho} = \partial_\rho V_{rad}~.
\ee
This tunnel configuration is relevant for temperatures that are below
the inverse bubble radius~\cite{Linde:1980tt,Nardini:2007me}. In the
present model the bounce solution results to be in this regime and the
quantum tunneling competes with thermal tunneling. In the conformal
case, the solution to the bounce equation is given by
\be
\mu(\rho) = \frac{2 \mu_0}{2 + \bar \kappa \rho^2 \mu_0^2 }~,
\ee
which leads to the action
\be
\label{eq:S4_near}
S_4 = 12 (Ml)^3 \frac{2 \pi^2}{3 \bar \kappa}~,
\ee
while the corresponding temperature vanishes. Small deviations from
the conformal case will only lead to a relatively small temperature
while sizeable temperatures can only be obtained when the conformal
symmetry is broken by a release point which is not too far away from the
minimum of the potential. In this regime our nearly conformal
approximation (\ref{eq:S4_near}) underestimates the real action that
could be reliably obtained in the thin-wall approximation.  Still one
can generally conclude that for the same release point the thermal
tunneling leads to larger temperatures than the quantum tunneling 
since the friction term in the bounce equation is smaller. Hence if
both tunneling modes are feasible the system tends to tunnel by thermal
fluctuations.

In the near conformal case a good approximation is given by using a
conformal potential normalized at the release point
\bea
\label{eq:pot_nconf}
V_{near-conf} &=& V_{rad}(\mu_0) \lp\frac{\mu}{\mu_0}\rp^4~,
\eea
such that
\be
\bar \kappa = -\frac{ V_{rad}(\mu_0)}{\mu_0^4}~.
\ee
\begin{figure}[b]
\includegraphics[width=0.95 \textwidth, clip ]{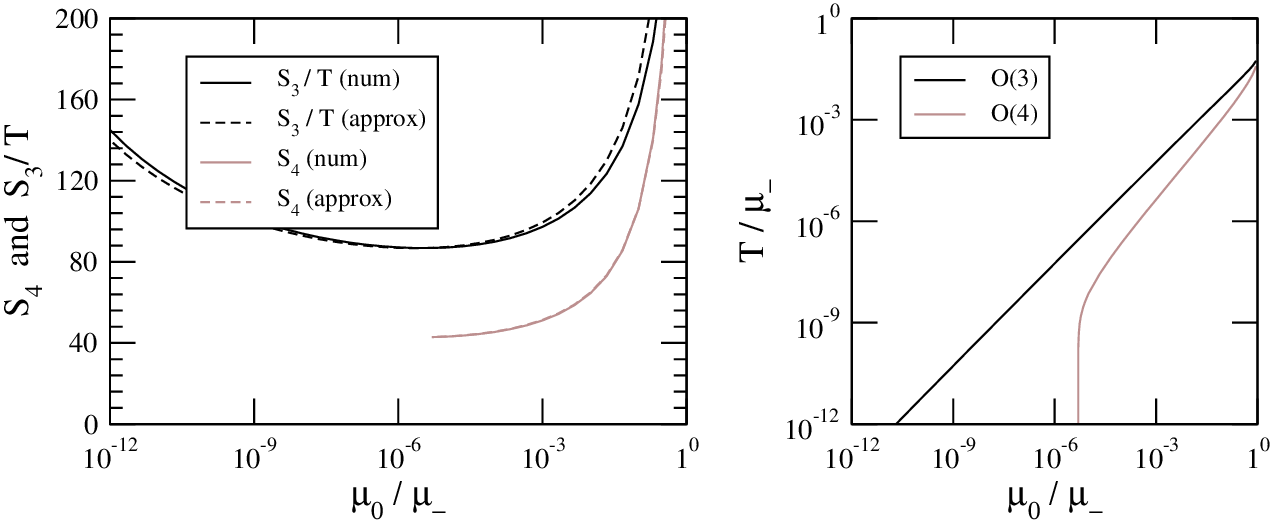}
\caption{
\label{fig:nconf}
The left plot shows the comparison of $S_3/T$ and $S_4$ as a function
of the release point $\mu_0$ between the approximation in
(\ref{eq:pot_nconf}) and the full numerical results. The two curves
for $S_4$ lie on top of each other and significant deviations only
occur for a release point very close to the minimum of the potential.
The right plot shows the temperature $T$ as a function of
the release point $\mu_0$. The used values are $v_1=4\, M^{3/2}$,
$v_2=v_1/3$, $N=3$, $\xi_-=1.5$ and $k_-l\simeq-0.019$.}
\end{figure}
In this approximation the minimum of the bounce action as a function
of $\mu_0$ (or equivalently $T$) is given by the maximum of
$V_{rad}(\mu_0) / \mu_0^4$. According to (\ref{eq:fitV}) the function
$V_{rad}(\mu)/\mu^4$ is nearly polynomial in terms of $\xi$ with one
extremum between $\xi_-$ and $\xi_+$. The derivative of this function
at the endpoints is given by
\be
d_\pm = \left. \partial_\xi \frac{V_{rad}(\mu)}{\mu^4} \right|_{\xi=\xi_\pm} 
= \left. -\frac{4}{l k_- \xi} \frac{V_{rad}(\mu)}{\mu^4} \right|_{\xi=\xi_\pm}
= \frac16 \frac{v^2_2}{M^3} (l k_+) (\xi_\pm-1)~.
\ee
A reasonable estimate of the position of the minimum is
\be
\frac{d_- \xi_- - d_+ \xi_+}{d_- - d_+}
= \xi_- + \xi_+ -1~,
\ee
with the approximate value
\be
\label{eq:kapp_in_min}
\bar \kappa_{max} = -\frac{d_- d_+}{d_- - d_+} \frac{\xi_- - \xi_+}{2}
= \frac1{12} \frac{v^2_2}{M^3} (l k_+) (\xi_- - 1)(1 - \xi_+)~.
\ee
For example a nearly maximal value is given for $\xi_- \approx 1.5$,
$v_1/v_2 \approx 3$. In this case the coefficient is given by $\bar
\kappa \approx 9.2 \times 10^{-3}\, v_1^2/M^3$.  For this choice of
parameters and with $v_1=4\, M^{3/2}$, $N=3$, $k_-l\simeq-0.019$,
which create the correct hierarchy, a comparison between this
approximation and the full numerical result is shown in
fig.~\ref{fig:nconf}. It turns out that the estimate is quite reliable
and in particular no appreciable difference is found for $S_4$.

The bulk cosmological constant is related to the five-dimensional
Planck mass by the flux $N$ of the background fields which, according
to the AdS/CFT correspondence, translates into the rank of the gauge
group of the CFT in four dimensions. In the present context this
relation is only known to leading order in $N$ and here we use
$(Ml)^3 = (N^2 -1)/16\pi^2$ as definition~\cite{Gubser:1999vj}.  Then
considering the maximal allowed $\bar \kappa$ for the thermal tunneling
action one obtains the bound
\be
S_3/T \gtrsim 61.5 \times (N^2-1) \lp \frac{v_1^2}{M^3} \rp^{-3/4}~,
\ee
and a nucleation temperature that is substantially smaller than the
scale $\mu_-$. Using the criterion (\ref{eq:S3_est}) for tunneling
this translates into
\be
\label{eq:bound_t}
N^2 - 1 \lesssim 2.3 \times \lp \frac{v_1^2}{M^3} \rp^{3/4}
\quad ({\rm thermal})\,,
\ee
which is quite similar to the bound obtained
in~\cite{Creminelli:2001th}. Analogously for the case of quantum
tunneling one finds
\be
S_4 \gtrsim 54.3 \times (N^2-1) \lp \frac{v_1^2}{M^3} \rp^{-1} ~,
\ee
and
\be
\label{eq:bound_q}
N^2 - 1 \lesssim 2.6 \times \lp \frac{v_1^2}{M^3} \rp
\quad ({\rm quantum})\,.
\ee
Depending on the value of $v_1^2/M^3$ the weakest bound
between~(\ref{eq:bound_t}) and (\ref{eq:bound_q}) constitutes the
minimal requirement on $N$ allowing for a first-order transition.

Alternatively one can saturate the bound of applicability of our
analysis (\ref{eq:vel_bound}) which reads, for $\xi_-$ not too close to
unity, as
\be
\frac{v_2^2}{M^3} \lp \xi_- - 1 - \frac{k_-}{k_+}\rp^2 \lesssim \frac{3}{2l^2}~,
\ee
such that 
\be
\label{eq:kapp_ub}
\bar \kappa \lesssim \frac12 \frac{(\xi_- -1)(1 - \xi_+ )}
{(\xi_- - 1 - k_-/k_+)^2}
\lesssim \frac12 ~,
\ee
which gives for thermal tunneling
\be
N \lesssim 6.8 \quad (\textrm{thermal})\,,
\ee
while for quantum tunneling one finds
\be
N \lesssim 12.5 \quad (\textrm{quantum})\,.
\ee
Thus there is a small window where quantum tunneling is possible while
thermal tunneling is not.

We conclude that even though the limits on tunneling derived including
backreactions are parametrically similar to the results obtained
without them our analysis shows that the constraint on the parameter
space used in the literature, $\phi^2 \ll
M^3$~\cite{Goldberger:1999uk, Creminelli:2001th, Randall:2006py,
  Kaplan:2006yi,Nardini:2007me}, is actually more conservative than
necessary.

\section{Gravitational wave observations\label{sec:gwo}}

The main input parameter for the determination of the gravitational wave
spectrum is the inverse duration of the phase transition that,
normalized to the Hubble parameter, is given by
\be
\beta / \sqrt{\Lambda} = T \frac{d}{dT} \lp \frac{S_3}{T}\rp~.
\ee
This equation assumes that the temperature is proportional to the
inverse scale factor, $T \propto \exp(-\sqrt{\Lambda} t)$, which is
also true in the AdS-S phase~\cite{Gubser:1999vj}.  In this section we
will present analytic as well as numerical estimates for
$\beta/\sqrt{\Lambda}$.

The dependence of the coefficient $\bar \kappa$ in the near-conformal case
is approximately polynomial at the release point $\xi_0$.  Besides
the tunnel action should grow to infinity if $\xi_0$ approaches either
$\xi_-$ or $\xi_+ \approx 2 - \xi_-$. A reasonable fit is then given by
\be
\bar \kappa(\xi_0) = 
\bar \kappa_{max} \lp 1 - \lp \frac{\xi_0 -1 }{\xi_- -1} \rp^2  \rp~.
\ee
Hereby and depending on the model parameters the value of
$\bar\kappa_{max}$ is somewhere in the range
\bea
\label{eq:kappa_limits}
3.1 \times 10^{-3} (N^2-1)^{4/3} < &\bar \kappa_{max}& < 0.56 \quad ({\rm thermal})\,, \nn \\
3.6 \times 10^{-3} (N^2-1) < &\bar \kappa_{max}& < 0.56 \quad ({\rm quantum})\,,
\eea
where the lower bound arises from the constraint that the system will
tunnel [eqs.~(\ref{eq:tunnel_num}) and (\ref{eq:S4_near})] while the
upper bound results from a numerical equivalent of (\ref{eq:kapp_ub})
obtained by exploring the parameter space constrained by the
requirement of small backreaction (\ref{eq:vel_bound}).

For the thermal tunneling action (\ref{eq:tunnel_num}) the inverse
duration of the phase transition is given by (notice that
approximately $\mu_0 \propto T$ as shown in fig.~\ref{fig:nconf})
\bea
\beta / \sqrt{\Lambda} &=& \frac34 \frac{S_3}{T} (l k_-) 
\frac{\xi_0}{\bar\kappa} \frac{d\bar\kappa}{d\xi_0} 
= \frac32 \frac{S_3}{T} (-l k_-) 
\frac{\bar \kappa_{max}}{\bar\kappa} \frac{\xi_0}{\xi_- -1} 
\frac{\xi_0 -1}{\xi_- -1} \nn \\
&=& \frac32 \frac{S_3}{T} (-l k_-) 
 \frac{\xi_0}{\xi_- -1} \frac{\bar \kappa_{max}}{\bar\kappa}
\sqrt{1- \frac{\bar \kappa}{\bar\kappa_{max}}}~.
\eea
At the first sight it seems that $\beta /\sqrt{\Lambda}$ can become arbitrarily large for $\xi_- \to
1$. However if $\xi_-$ approaches the unity the constraint
(\ref{eq:vel_bound2}) becomes more severe than (\ref{eq:vel_bound})
and $\bar \kappa_{max}$ is according to (\ref{eq:kapp_in_min}) bounded
by
\be
\bar \kappa_{max} \simeq \frac1{12} \frac{v^2_2}{M^3} (l k_+) (\xi_- - 1)(1 - \xi_+)
< \frac2{l k_-} \frac{v^2_2}{v_1^2} (\xi_- - 1)(1 - \xi_+)~,
\ee
such that the maximum possible value of $\beta / \sqrt{\Lambda}$ in
fact becomes constant in this limit. Hence one can obtain a
conservative estimate by assuming that $\xi_-$ is in the transition
region where the two constraints (\ref{eq:vel_bound2}) and
(\ref{eq:vel_bound}) are equally severe
\be
\xi_- - 1 \simeq \sqrt{-l k_-} \frac12  \frac{v_1}{v_2}~,
\ee
and hence
\be
\beta / \sqrt{\Lambda} < 
3 \sqrt{ - l k_-} \, \frac{S_3}{T} \, \frac{v_2}{v_1} 
\frac{\bar \kappa_{max}}{\bar\kappa}
\sqrt{1- \frac{\bar \kappa}{\bar\kappa_{max}}}~.
\ee
Noting that
\be
\sqrt{-l k_-} \, \frac{v_2}{v_1} \simeq \sqrt{-l k_-} \, e^{k_- r_-}
\lesssim \sqrt{\frac{l}{2 \, e \, r_-}}~,
\ee
one obtains the bound
\be
\label{eq:beta_t}
\beta / \sqrt{\Lambda} \lesssim \sqrt{\frac{9 \, l}{2 \, e \, r_-}} \, \frac{S_3}{T} 
\frac{\bar \kappa_{max}}{\bar\kappa}
\sqrt{1- \frac{\bar \kappa}{\bar\kappa_{max}}}
\approx 30 \frac{\bar \kappa_{max}}{\bar\kappa}
\sqrt{1- \frac{\bar \kappa}{\bar\kappa_{max}}}~.
\ee
Eq.~(\ref{eq:kappa_limits}) for the thermal case implies
\be
\frac{\bar \kappa_{max}}{\bar \kappa} \lesssim \frac{180}{(N^2-1)^{4/3}}~,
\ee
which finally yields a bound on $\beta/\sqrt{\Lambda}$\,.

Analogously for quantum tunneling one finds
\be
\label{eq:beta_q}
\beta / \sqrt{\Lambda} \lesssim \sqrt{\frac{8\, l}{ e \, r_-}} \, S_4 
\, \frac{\bar \kappa_{max}}{\bar\kappa}
\sqrt{1- \frac{\bar \kappa}{\bar\kappa_{max}}}
\lp \frac{T}{\mu_0} \frac{d\mu_0}{dT} \rp~.
\ee
The last factor is typically smaller than in thermal tunneling (where
it is unity,~\cf fig.~\ref{fig:nconf}) so that commonly quantum
tunneling happens only when the system cannot tunnel by thermal
fluctuations, which implies
\be
\bar \kappa_{max} < \lp \frac{140}{290} 16 \pi^2 \rp^{-4/3} (N^2-1)^{4/3}~.
\ee
Besides the largest $\bar\kappa$ that can be realized in the present
model is generally given by $\bar \kappa_{max} < 0.56$ according to
eq.~(\ref{eq:kappa_limits}). On the other hand the quantum tunneling
condition (\ref{eq:kappa_limits}) implies
\be
\bar \kappa > \lp 2 \times 140 \rp^{-1} (N^2-1)~,
\ee
such that successful quantum transitions which are not spoiled by
earlier thermal tunneling lead to the constraint
\be
\frac{\bar \kappa_{max}}{\bar \kappa} \lesssim \textrm{min} \, 
\left[ \, 0.85\,  (N^2 -1)^{1/3}, \,
155 \, (N^2 -1)^{-1}\right] \,.
\ee
The first bound increases with $N$ and arises from the requirement
of no thermal tunneling while the second bound stems
from viable quantum tunneling and decreases with $N$. As a result, as
far as $N$ decreases the quantum tunneling bounds on
$\beta/\sqrt{\Lambda}$ become first weaker (while thermal tunneling
becomes less likely) and then stronger (as thermal tunneling is in
general not possible for $N>7$). Finally for $N>12$ tunneling is
impossible altogether.

The approximate values and numerical results on the bound on
$\beta/\sqrt{\Lambda}$ are given in Table~\ref{tab:pt}.  Already for
$N=2$ one observes a rather strong phase transition ($\alpha\gtrsim
1$).  We see that for small $N$, when the release point is close to
the minimum of the potential, our approximation overestimates the
numerical result for $\beta/\sqrt{\Lambda}$, while for larger values
of $N$ it yields fairly precise results.

The second ingredient for the gravity wave spectrum is the vacuum
energy normalized to the radiation energy of the system (traditionally
denoted by $\alpha$ in the literature) 
\be
\alpha = \frac{12(M l)^3(V_{rad}(\infty) - V_{rad}(r_-))}{\rho_{radiation}} 
= \frac{l^{-2} \Lambda}{2 \pi^4 T^4_n}~.
\ee
In the present system this parameter is much larger than unity as it
is shown in Table~\ref{tab:pt} due to the large supercooling. This
\begin{table}[htp]
\begin{tabular}{|c||c|c|c|| c| c |}
\hline
& \multicolumn{3}{|c|}{thermal} & \multicolumn{2}{|c|}{quantum} \\
\hline
\hline
$N$ & \quad $\beta_{approx}/\sqrt{\Lambda}$ \quad  & \quad $\beta_{num}/\sqrt{\Lambda}$ \quad  
& $\alpha_{num}$ & 
$\quad\beta_{num}/\sqrt{\Lambda}\quad$ & $\alpha_{num}$ \\
\hline
\hline
2 & 	$<1230$ & 	$<770$ & 	$>3.2$  	& 
$<15$  & $>1.0 \times 10^{11}$ \\
\hline
3 & 	$<235$ & 	$<315$ & 	$>10$  	& 
$<33$  & $>5 \times 10^7$ \\
\hline
4 & 	$<131$ & 	$<143$ & 	$>50$ 		& 
$<45$ & $>2.5 \times 10^6$ \\
\hline
5 & 	$<62$ & 	$<67$ & 	$\quad>800\quad$ 		& 
$<56$  &  $>4 \times 10^5$ \\
\hline
6 & 	$<29$ & 	$<30$ & 	$>10^5$ 	& 
$<63$   &  $>1.3 \times 10^5$\\
\hline
7 & 	$<6.5$ & 		$<6.0$ &  	
$>10^8$ \quad 		& 
$<71$  &  $> 5 \times 10^4$  	\\
\hline
8 &\multicolumn{3}{|c|}{}& $<54$ & $>5 \times 10^5$ \\
\hline
9 &\multicolumn{3}{|c|}{}& $<40$  & $>8 \times 10^6$\\
\hline
10 &\multicolumn{3}{|c|}{}& $<27$ & $>3 \times 10^8$\\
\hline
11 &\multicolumn{3}{|c|}{}& $<17$ & $>4 \times 10^{10}$ \\
\hline
$\quad 12\quad$ &\multicolumn{3}{|c|}{}& $<5.4$ & $\quad>5 \times 10^{14}\quad$\\
\hline
\end{tabular}
\caption{
\label{tab:pt} 
Upper limits on $\beta/\sqrt{\Lambda}$ and lower limits on $\alpha$ for
all possible values of $N$.  
}
\end{table}
also implies that the nucleated bubbles expand with near luminal
velocities. The vacuum energy is very efficiently transformed into
bulk motion of the plasma or directly into kinetic energy of the
bubble wall and resides in a very thin shell around the bubble
wall~\cite{EKNS}. In this sense the phase transition is extremely
strong.

The main mechanisms of gravity wave production during a first-order
phase transition are bubble collisions~\cite{Kosowsky:1991ua,
Kosowsky:1992rz, Kosowsky:1992vn, Kamionkowski:1993fg, Caprini:2007xq,
Huber:2008hg}, turbulence~\cite{Kosowsky:2001xp, Dolgov:2002ra,
Gogoberidze:2007an, Caprini:2009fx} and magneto-hydrodynamic
turbulence~\cite{Caprini:2006jb, Caprini:2009yp}. In our analysis we
will focus on the production mechanism by bubble collisions for the
following reasons. First, while reliable information about the gravity
wave spectrum in the case of bubble collisions is provided by computer
simulations, the analysis of turbulence relies typically on additional
assumptions (\eg on the overall normalization of the
spectrum). Second, the peak frequency of the spectrum produced by
turbulence compared to the one by bubble collisions is typically
suppressed by the eddy velocity that occurs in the turbulent
fluid motion~\cite{Kosowsky:2001xp, Dolgov:2002ra}. At the same time
the spectrum for turbulence falls off faster than the one by bubble
collisions. Hence even if the different components are of similar size
they might be disentangled because they lead to a double peak
structure or at least a knee in the spectrum. Finally, recent studies
have shown that very strong phase transitions most probably have a
runway behavior of the wall~\cite{Bodeker:2009qy} and in this case
most of the vacuum energy is transformed into gradient/kinetic energy
of the Higgs and not into collective bulk motion of the
plasma~\cite{EKNS}. This should also reduce the portion of energy that
leads to turbulent plasma motion.  In the following we will focus on
the spectrum produced by bubble collisions.

In the above discussed limit, $\alpha \gg 1$, the energy fraction for
colliding bubbles of the gravitational radiation at the time of
production is given by
\cite{Huber:2008hg}
\be
h^2 \Omega^* = 7.7 \times 10^{-2} \lp \frac{\sqrt{\Lambda}}{\beta} \rp^2~.
\ee
In a standard cosmology this energy density is diluted to today's
observed energy fraction
\be
h^2 \, \bar \Omega = 1.3 \times 10^{-6} 
\lp \frac{\sqrt{\Lambda}}{\beta} \rp^2~.
\ee
The only assumptions that enters here is that the Universe is
dominated by radiation after the phase transition. Immediately after
the phase transition the energy fraction stays constant up to the time
of matter-radiation equality when the energy fraction of gravitational
radiation starts to become suppressed together with all the other
light components of the plasma.

\begin{figure}[b]
\includegraphics[width=0.65\textwidth, clip ]{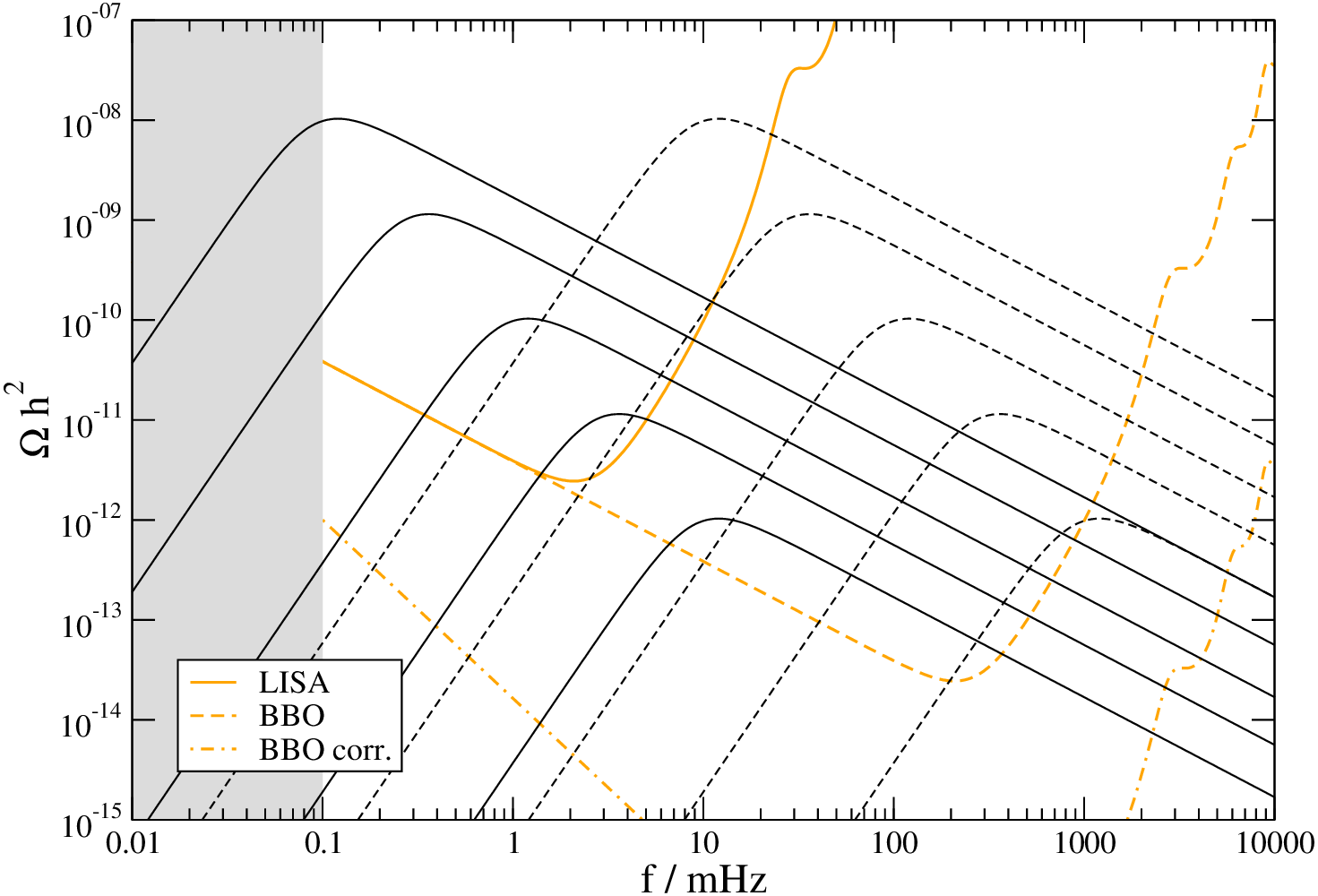}
\caption{\label{fig:shape} 
Several example spectra of gravitational waves. The straight (dashed)
lines are for a reheating temperature $T_{reh}/\sqrt{\Lambda}=10^{16}$
($T_{reh}/\sqrt{\Lambda}=10^{14}$). From bottom to top the plots use
$\beta/\sqrt{\Lambda} = \left\{ 1000,300,100,30,10\right\}$. The
sensitivities of the LISA and BBO experiments are taken
from~\cite{Buonanno:2004tp}.  }
\end{figure}

The frequency peak of the spectrum at time of generation is in the
limit of very strong phase transition, $\alpha \gg 1$, and large wall
velocities, $v_b \sim 1$, given by
\cite{Huber:2008hg}
\be
f^* = 0.23\, \beta~.
\ee
The red-shift to today's observed spectrum depends on the reheating
temperature after the phase transition according to 
\be
\bar f = 0.23 \, \beta \,
\frac{T_0}{T_{reh}}~,
\ee
where $T_0$ is the observed temperature of the cosmic microwave
background. The spectrum is of
form~\cite{Huber:2008hg}
\be
\Omega(f)  = \bar \Omega 
\frac{3.8 \,  f^{2.8} \bar f^{1.0}}
{1.0 \bar f^{3.8} + 2.8 f^{3.8}}~. 
\ee
Several sample spectra are given in Fig.~\ref{fig:shape}.

As long as the expansion is negligible during the phase transition the
reheating temperature $T_{reh}$ can be determined by energy
conservation. Since the phase transition is extremely supercooled at
the typical nucleation temperatures the energy density of AdS-S phase
is comparable to the one of pure AdS and thus
\be
\label{eq:Treh}
g_* \frac{\pi^2}{30} T^4_{reh} \simeq  12 (Ml)^3 (V_{rad} (0) - V_{rad} (\mu_-)) =
6 \, (Ml)^3 \, l^{-2} \Lambda~, 
\ee
where $\Lambda=\Lambda_\infty$ and $g_*$ denotes the effective number
of degrees of freedom just after the phase transition. Notice that for
$g_* \lesssim 120 \pi^2(Ml)^3$ the critical temperature, given by
\be
4 \pi^4 T_c^4 = 6 l^{-2} \Lambda~,
\ee
is smaller than the reheating temperature and percolation is followed
by a period of phase coexistence~\footnote{Assuming that some SM
fields be composite relaxes the condition on $g_*$.}. Nevertheless
we will assume that the red-shift of the gravity waves is given by the
naive expression involving the reheating temperature (\ref{eq:Treh}).

We now give a conservative estimate for the range of possible values
for $T_{reh}/\sqrt{\Lambda}$ in a realistic model. According to
(\ref{eq:lambda_inf}) and (\ref{eq:m_approx}) the expansion parameter
is related to the radion mass by
\be
\Lambda \approx \frac18 m_{rad}^2 \, e^{-2r_-/l}~,
\ee
such that together with the definition of the four-dimensional Planck
mass, $(M_P l)^2 = (Ml)^3$, one obtains
\be
g_* \frac{\pi^2}{30}\frac{T^4_{reh}}{\Lambda^2} 
= 48 \frac{M^2_P}{m_{rad}^2} e^{2r/l}~.
\ee
Introducing the TeV scale $\mu_- = l^{-1} e^{-r_-/l}$ and the relation
$(M_P l)^2=(N^2-1)/16 \pi^2$ the peak frequency can be written as
\be
\bar f = 1.77 \times 10^{-3} \textrm{ mHz } \frac{\beta}{\sqrt{\Lambda}}
(N^2-1)^{1/4}
\sqrt{\frac{m_{rad} \, \mu_-}{\textrm{TeV}^2}}~.
\ee
Realistically the last factor varies within a range of $[0.3, 10]$.
\begin{figure}[t]
\includegraphics[width=0.65\textwidth, clip ]{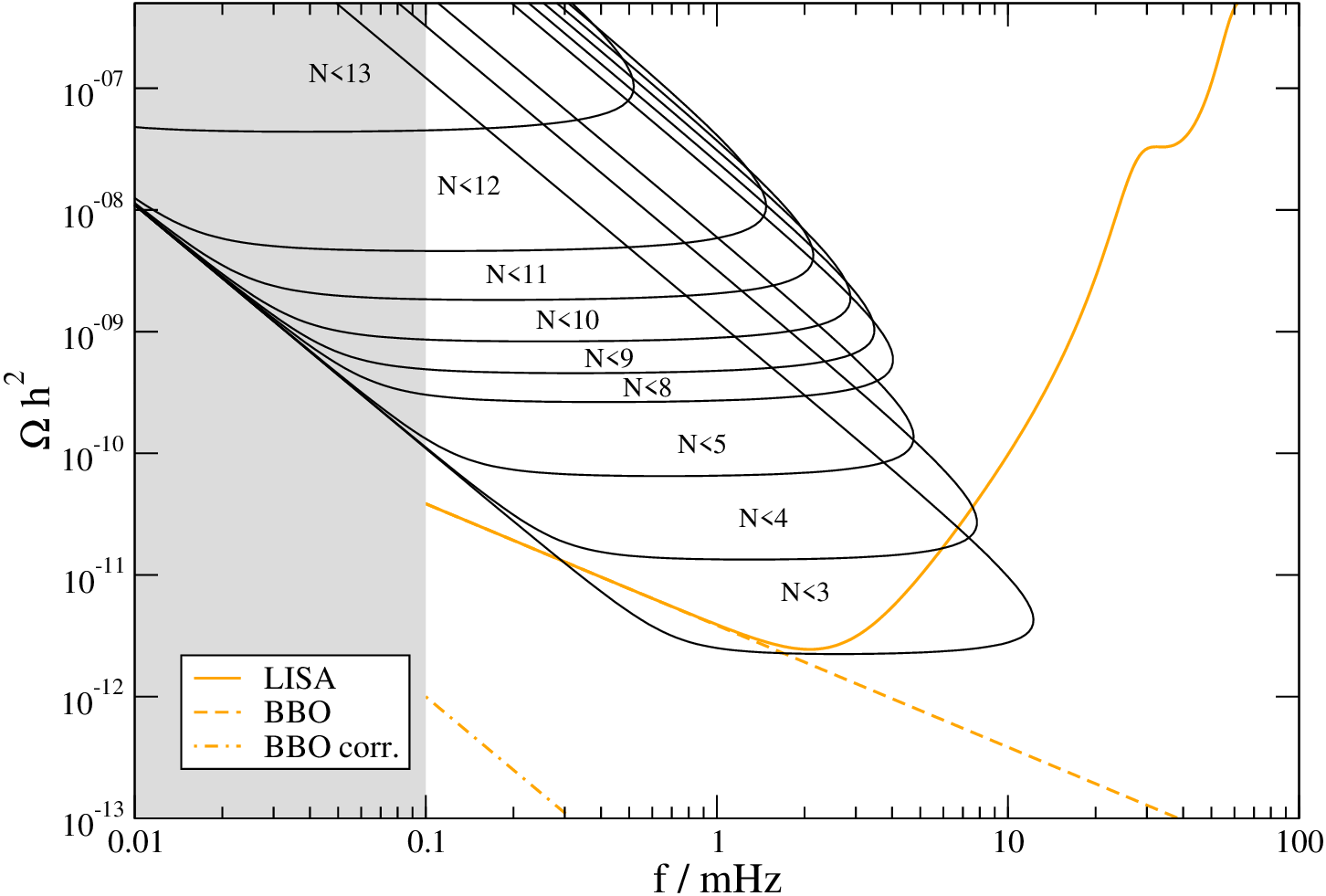}
\caption{\label{fig:spec} The regions denote the possible positions
  for the peak of the gravity wave spectrum depending on the parameter
  $N$. The signal will be detected by LISA, BBO or BBO correlated when
  it stands above their respective lines.}
\end{figure}
The corresponding ranges for the possible positions of the peak
structure of the gravitational wave spectrum compared with
sensitivities of future experiments are shown in
Fig.~\ref{fig:spec}. For $N\leq3$ the phase transition could be weak
enough to prohibit detection by LISA. Besides for these values the
$1/N$ expansion is clearly questionable and stringy loop contributions
invalidate our analysis. On the other hand for $N\geq4$ the model
would lead to a gravitational wave signal that is observable by LISA.

\section{Conclusion\label{sec:concl}}

Motivated by the sizeable impact of backreactions on the radion mass
determined with the so-called superpotential
method~\cite{DeWolfe:1999cp,Csaki:2000zn}, we studied backreactions to
the radion potential in a perturbative scheme. Our approach includes
backreactions on the five-dimensional gravitational fields from the
bulk scalar under the constraints in eqs.~(\ref{eq:vel_bound2}) and
(\ref{eq:vel_bound}) which are less severe than the usual assumption
of weak fields, $\phi^2 \ll M^3$~\cite{Goldberger:1999uk,
  Creminelli:2001th, Randall:2006py, Kaplan:2006yi,
  Nardini:2007me}. An immediate consequence of the backreactions is
the necessity to generalize the metric to include the Hubble expansion
of space-time in four dimensions. In our framework a judicious choice
of brane tensions allows to reproduce the radion potential found by
Goldberger and Wise as well as (within the applicability of our
approach) the radion mass obtained with the superpotential method. We
then clarify the apparent paradox that the radion mass computed in the
GW and superpotential frameworks scale (for a bulk scalar mass $m^2
\sim 4 k_-/l$) as $|k_-|^{3/2}$ and $|k_-|^2$, respectively. On top of
that in the regime of sizeable detuning of brane tensions we found
that a scaling proportional to $k_-$ is possible. This results in a
larger radion mass for fixed TeV scale and in turn it leads to a
deeper radion potential which has significant impact on the thermal
phase transition of the system.

The phase transition constitutes a serious problem for the holographic
interpretation of models with a radion stabilized by a bulk
scalar~\cite{Creminelli:2001th, Randall:2006py, Kaplan:2006yi,
  Nardini:2007me, Reece:2010xj}: the tunnel action scales as $N^2$
with the flux $N$ of the background field such that the metastable
symmetric phase becomes, for large values of $N$, effectively stable and
the conformal symmetry is never broken during the course of the
Universe. One possible solution to this problem is to consider more
general scalar potentials (as \eg done in
ref.~\cite{Hassanain:2007js}) which modify the behavior of the bulk
fields considerably in the IR. Analogously large backreactions can
have potentially the same effect by decreasing the 'effective' $N$
through a stronger warping close to the TeV brane.

In our analysis of the phase transition, and compared to former
work~\cite{Creminelli:2001th, Randall:2006py}, we did not use the
common thick or thin wall approximations to obtain the tunnel action
but devised an approximation that is tailor-made for nearly-conformal
potentials. Besides notice that we used the kinetic term obtained by
decoupling the Einstein equations~\cite{Csaki:2000zn} which differs by
a factor of $2$ from the one used in~\cite{Creminelli:2001th,
  Randall:2006py, Nardini:2007me}. The main consequence of our
approach is that the tunneling action depends logarithmically on the
temperature and can lead to a couple of e-folds of low scale inflation
without tuning the model parameters (this was already observed in the
numerical analysis of ref.~\cite{Nardini:2007me}). While a few e-folds
of inflation cannot solve the horizon problem a low phase transition
temperature has a large impact on gravitational wave production since,
in this case, the energy stored in the vacuum bubbles during
percolation at the end of the phase transition is many orders of
magnitude larger than the energy stored in the thermal plasma.

In summary we found that, in the regime of large
backreactions, the deeper radion potential leads to a significantly
weaker phase transition and numerically the absolute limit $N<13$ applies.
Besides we reanalyzed the gravitational wave spectrum produced by the
first-order phase transition. We conclude that as long as stringy
corrections can be neglected (specifically $N>3$), the model leads to
a stochastic background of gravitational radiation that can be
observed by LISA (see fig.~\ref{fig:spec}).

\section*{\large Acknowledgments}
Work supported in part by the European Commission under the European
Union through the Marie Curie Research and Training Network
``UniverseNet'' (MRTN-CT-2006-035863); by the Spanish
Consolider-Ingenio 2010 Programme CPAN (CSD2007-00042); by CICYT,
Spain, under contract FPA 2008-01430; by the Belgian IISN convention
4.4514.08 of FNRS; and by ERC starting grant Cosmo@LHC.

\appendix

\section{Linearized system of equations and the radion mass\label{sec:appx}}

In order to calculate the radion mass we will follow the analysis of
ref.~\cite{Csaki:2000zn} including the expansion of the
Universe. For the perturbations of the metric we use the Ansatz
\bea
\phi(x,r) &=& \phi_0(r) + \varphi(x,r)~,  \nn \\
\label{eq:metric_pert}
ds^2 &=& e^{2A+2F(x,r)} \bar g_{\mu\nu} dx^\mu dx^\nu -
(1 - 2 F(x,r)) ^2 dr^2~,
\eea
where $\bar g$ denotes the induced $dS_4$ metric as in eq.~(\ref{eq:def_g}) 
\be
\bar g_{\mu\nu} dx^\mu dx^\nu = dt^2- 
e^{2 \sqrt{\Lambda} t} (dx_1^2 + dx_2^2 + dx_3^2 )~,
\ee
and solve the linearized Einstein equations, $\delta
R_{\mu\nu} = \delta T_{\mu\nu} $. The corresponding entries for the
Riemann tensor are given by 
\bea
\delta R_{\mu\nu} &=& 
  \bar g_{\mu\nu}  
\left( -\square F + F'' + 10 A' F' + 6 A'' F + 24 {A'}^2 F \right)~, \nn \\
\delta R_{\mu5} &=& 
 - 6A' \partial_\mu F - 3 \partial_\mu F^\prime~, \nn \\
\delta R_{55} &=& 
-2 \square F - 16 A' F' - 4 F'' ~, 
\eea
where we defined the d'Alembertian operator in curved space time ($\bar g
= \det \bar g_{\mu\nu}$)
\be
\square F = \bar g^{-\frac12} \partial_\mu
(\bar g^{\frac12} \bar g^{\mu\nu} \partial_\nu F)
= \bar g^{\mu\nu} \partial_\mu \partial_\nu F 
+ 3 \sqrt{\Lambda} e^{-2A}\partial_t F ~ .
\ee
The energy-momentum tensor is given by
\bea
\delta T_{\mu\nu} &=&
- \frac43 \bar g_{\mu\nu} \left( V'(\phi_0) \varphi + 2 V(\phi_0) F \right) \nn \\
&& -\frac23 \bar g_{\mu\nu} \sum_i \left(
\frac{\partial \lambda_i(\phi_0)}{\partial \phi} \varphi
+4 \lambda_i(\phi_0) F \right) \delta(r-r_i)~, \nn \\
\delta T_{\mu5} &=& 
2 \phi_0' \partial_\mu \varphi ~,\nn \\
\delta T_{55} &=&
4 \phi_0' \varphi' + \frac43 V'(\phi_0) \varphi - \frac{16}3 V(\phi_0) F \nn \\
&& + \frac83 \sum_i \left(
\frac{\partial \lambda_i(\phi_0)}{\partial \phi} \varphi
 -2 \lambda_i(\phi_0) F \right) \delta(r-r_i)~, 
\eea
and the scalar equation is given by
\bea
\square \varphi - \varphi'' - 4 A' \varphi' 
+ \frac{\partial^2 V}{d\phi^2} (\phi_0) \varphi  &=& \nn \\
&& \hskip -5 cm
- \sum_i \left( \frac{\partial \lambda_i(\phi_0)}{\partial \phi} \varphi
 -2 \lambda_i(\phi_0) F \right) \delta(r-r_i)
- 6 \phi_0' F' - 4 \frac{\partial V}{\partial \phi} F~.
\eea
The equation for $R_{\mu5}$ can be integrated to yield
\be
\label{eq:int_mu5}
\phi_0' \varphi = - \frac32 (F' + 2A' F)~.
\ee
Next consider the equation $\frac14 \bar g^{\mu\nu} \delta R_{\mu\nu} +
\delta R_{55}$ in the bulk
\be
3 \square F + 6 A' F' 
- 24 A'^2 F - 6 A'' F + 3 F'' =
 8 V F - 4 \phi^\prime_0 \varphi'~.
\ee
With eqs.~(\ref{eq:int_mu5}) 
\be
-\frac23 \phi_0' \varphi' =  F'' + 2A'' F + 2 A' F'
-\frac{\phi_0''}{\phi_0'}(F' + 2A' F)~,
\ee
and eqs.~(\ref{eq:Einstein})-(\ref{eq:Einstein2}) 
\be
- \frac43 V = A'' + 4 A'^2 - 3 \Lambda e^{-2A}~,
\ee
this can be brought into the form 
\bea
\square F &=& F'' + 2 A' F' + 4 A'' F  - 2 \frac{\phi_0''}{\phi_0'} 
\left[ F' + 2 A' F \right] \nn \\
&&  + 6 \Lambda e^{-2A} F~.
\eea
A minimally coupled scalar in four-dimensional curved space time
fulfills the equation 
\be
(e^{2A} \square +  m^2) F = 0~,
\ee
which leads to the equation
\be
(- m^2 - 6 \Lambda) e^{-2A} F = F'' + 2 A' F' + 4 A'' F  - 2 \frac{\phi_0''}{\phi_0'} 
\left[ F' + 2 A' F \right]~.
\ee
Hence in an
expanding Universe the sole difference on the equation for the radion mass with respect to the usual radion equation can be taken into account by a shift in the mass parameter
\be
\hat m^2 = m^2 + 6 \Lambda~.
\ee
Since the Hubble parameter is many orders smaller than the radion
mass this shift can quite generally be neglected. Using the Ansatz
$F = e^{2A} \hat F$ this leads to 
\be
\label{eq:app_final}
\hat F'' + 2 A'' \hat F - 2 A' \hat F' - 
2 \frac{\phi_0''}{\phi_0'} \hat F' = - \hat m^2 e^{-2A} \hat F~,
\ee
which is the starting point of the analysis in section \ref{sec:rad_mass}.


\end{document}